\documentclass{aa}
\usepackage[utf8]{inputenc}

\usepackage{graphicx,ulem}
\usepackage{longtable}
\usepackage{booktabs}
\usepackage{float}
\usepackage{dcolumn}
\usepackage{graphics,epsfig}
\usepackage{amsmath,amssymb,latexsym,mathrsfs}
\usepackage[breaklinks, colorlinks, citecolor=blue]{hyperref}
\usepackage{bm}
\usepackage{color}
\usepackage{subfigure}
\usepackage{multirow}

\newcommand\morehorsp{\rule[-3mm]{0mm}{8mm}}

\begin{document}

\title{Mass bias evolution in tSZ cluster cosmology}
\author{Laura Salvati\inst{1}\inst{*} \and Marian Douspis\inst{1} \and Anna Ritz\inst{2} \and Nabila Aghanim\inst{1} \and Arif Babul\inst{3}}
\institute{\inst{1} Institut d'Astrophysique Spatiale, CNRS (UMR 8617) Université Paris-Sud, Bâtiment 121, Orsay, France \\
\inst{2} Magistère de Physique Fondamental, Université Paris Sud, Orsay, France \\
\inst{3} Department of Physics and Astronomy, University of Victoria, Victoria, BC V8P 1A1, Canada\\
\inst{*}\email{laura.salvati@ias.u-psud.fr}}
\authorrunning{Salvati et al.}
\date{}

\abstract{
Galaxy clusters observed through the thermal Sunyaev-Zeldovich (tSZ) effect are a recent cosmological probe. The precision on the cosmological constraints is affected mainly by the current knowledge of cluster physics, which enters the analysis through the scaling relations. Here we aim to study one of the most important sources of systematic uncertainties, the mass bias, $b$. We have analysed the effects of a mass-redshift dependence, adopting a power-law parametrisation.
We applied this parametrisation to the combination of tSZ number counts and power spectrum, finding a hint of redshift dependence that leads to a decreasing value of the mass bias for higher redshift. We tested the robustness of our results for different mass bias calibrations and a discrete redshift dependence. We find our results to be dependent on the clusters sample that we are considering, in particular obtaining an inverse (decreasing) redshift dependence when neglecting $z<0.2$ clusters. We analysed the effects of this parametrisation on the combination of cosmic microwave background (CMB) primary anisotropies and tSZ galaxy clusters. We find a preferred constant value of mass bias, having $(1-b) =0.62 \pm 0.05$. The corresponding value of $b$ is too high with respect to weak lensing and numerical simulations estimations. Therefore we conclude that this mass-redshift parametrisation does not help in solving the remaining discrepancy between CMB and tSZ clusters observations.
}

\maketitle

\section{Introduction} 

Galaxy clusters are able to track the recent evolution of the large scale structure, describing the matter density field. Therefore, they represent an important cosmological probe. In recent decades, different observations of cluster samples in X-rays \citep{Boehringer:2017wvr,Pacaud:2018zsh}, optical \citep{Rykoff:2016trm}, and millimetre wavelengths \citep{2014A&A...571A..20P,Ade:2015fva,deHaan:2016qvy,Bocquet:2018ukq} have substantially improved the constraints on cosmological parameters obtained from this probe.

In this paper we have considered galaxy clusters observed through the thermal Sunyaev-Zeldovich (tSZ hereafter) effect \citep{Sunyaev:1970er}. The latest cosmological analysis of tSZ galaxy clusters \citep{2014A&A...571A..20P,2016A&A...594A..22P,Ade:2015fva,deHaan:2016qvy} have shown that precision on the cosmological parameters is affected by systematic uncertainties, mainly related to cluster physics and theoretical assumptions -- see for example, the discussion in \citealt{McCarthy:2003pa,McCarthy:2003dr,Poole:2007vt,Hoekstra:2012ks,Mahdavi:2012zy,Ruan:2013osa,Sakr:2018new}. In particular, we have introduced uncertainties related to the calibration of the scaling relations between the survey observables and the real mass of the cluster. These relations are needed to extract cosmological information. A key element of these scaling relations is the mass bias. The bias arises from the assumption of hydrostatic equilibrium when estimating the cluster mass, and it is therefore related to the ratio between the estimated and the real mass of the cluster (cf. Section~\ref{sec:method}). In practice, it may also account for all possible observational biases, such as absolute instrument calibration, temperature inhomogeneities, and residual selection bias.
In order to quantify this bias, we use weak lensing (WL) mass reconstructions, which are supposed to provide an unbiased estimate of the true mass of the cluster, though with a larger scatter --  see for example the discussion in \citealt{Mahdavi:2007sd,Meneghetti:2009ti,Zhang:2010hma,Mahdavi:2012zy,Hoekstra:2015gda,Smith:2015qhs}.

In a previous paper \citep{Salvati:2017rsn} we show that the well known discrepancy on cosmological parameters, in particular on $\sigma_8$, obtained from cosmic microwave background radiation (CMB) primary anisotropies and tSZ observations \citep{2014A&A...571A..20P,2016A&A...594A..22P,Ade:2015fva} is substantially reduced, thanks to the lower value of the optical depth provided by \cite{Aghanim:2016yuo}. This result is also confirmed in 
the latest Planck release \citep{Aghanim:2018eyx}. 

Nevertheless, we show that a tension is still present on the value of the mass bias, when we compare the estimation from tSZ probes and baryon acoustic oscillations (BAO) and tSZ and CMB data combination.
This remaining discrepancy may be related to the general calibration of the adopted tSZ scaling relations. As described in \cite{2014A&A...571A..20P,Ade:2015fva}, these relations are calibrated through X-ray observations on a sub-sample of the total cosmological catalogue. They also provide an explicit mass-redshift evolution.
A complete re-calibration of these scaling relations is beyond the scope of this paper. We decided to focus only on the mass bias, since the correct evaluation of this quantity remains an open issue when using galaxy clusters as a cosmological probe.

In particular, in this paper we have analysed a possible variation for the mass bias, through an explicit mass-redshift parametrisation. Indeed, as shown for example in \cite{Smith:2015qhs}, the mass calibration changes when considering clusters samples in different redshift ranges.
We have investigated the effects of this parametrisaiton on the CMB-tSZ mass bias discrepancy. In order to further explain our results and their impact on tSZ cosmological constraints, we analysed the effects of a redshift-binned mass bias and different sample selections, on both redshift and signal-to-noise ratio. We used measurements from the Planck satellite for the galaxy cluster number counts \citep{Ade:2015fva}, in combination with the angular power spectrum of warm-hot gas, from Planck \citep{2016A&A...594A..22P} and the South Pole Telescope (SPT, \citealp{George:2014oba}), starting from the analysis done in \cite{Salvati:2017rsn}. 

The paper is organised as follows: in Section \ref{sec:method} we describe the approach we use in the analysis, presenting our results in Sections \ref{sec:results} and \ref{sec:tests}. We discuss the results and derive our final conclusions in Sections \ref{sec:discussion} and \ref{sec:conclusions}.

\section{Method}\label{sec:method}

In this analysis, we have studied the dependence of the mass bias with respect to mass and redshift, exploiting the combination of galaxy clusters number counts and power spectrum, following the approach described in \cite{Salvati:2017rsn}.
We have made use of the cluster sample provided by \cite{Ade:2015gva} (PSZ2 cosmo sample hereafter), consisting of 439 clusters obtained from the $65\%$ cleanest part of the sky, in redshift range $z=[0,1]$ and above the signal-to-noise ratio threshold of 6. Following the analysis done in \cite{Ade:2015fva}, we sampled on both redshift and signal-to-noise ratio. We also used the Planck estimation of tSZ power spectrum \cite{2016A&A...594A..22P}, in the redshift range $z = [0, 3]$ and in the mass range $M_{500} = \left[ 10^{13}h^{-1} M_{\odot}, 5\cdot 10^{15} h^{-1} M_{\odot} \right]$. We combined Planck data with the estimation from SPT at $\ell = 3000$ \cite{George:2014oba}.

\subsection{tSZ probes}

The intensity of the tSZ effect, in a given direction of the sky $\hat{\bf n}$, is measured through the thermal Compton parameter $y$
\begin{equation}\label{eq:y}
y(\hat{\bf n}) = \int n_{\text{e}} \dfrac{k_{\text{B}}T_{\text{e}}}{m_{\text{e}} c^2} \, \sigma_{\text{T}} \, ds \, ,
\end{equation}
with $k_{\text{B}}$ being the Boltzmann constant, $\sigma_{\text{T}}$ the Thomson scattering cross section, and $m_{\text{e}}$, $n_{\text{e}}$, and $T_{\text{e}}$ the electron mass, number density, and temperature, respectively.

We briefly recall here the definition of the tSZ observables and probes that we have used in this analysis. For an extensive description we refer the reader to our previous paper \citep{Salvati:2017rsn} and references therein.

Following the usual approach for tSZ observations, we define galaxy clusters properties within $\Delta_c = 500$. Therefore we considered cluster measurements within a sphere of radius $R_{500}$, within which the cluster mean mass over-density is 500 times the critical density at that redshift, $\rho_c(z)$. As a consequence, the cluster mass is defined as
\begin{equation}\label{eq:M500}
M_{500}= \dfrac{4 \pi}{3} R_{500}^3 500 \rho _{\text{c}} (z) \, .
\end{equation}
Following  \cite{Ade:2015gva}, we considered $Y_{500}$, which is the Compton $y$-profile integrated within a sphere of radius $R_{500}$, and the cluster angular size, $\theta _{500,}$ to be observables for cluster detection.
In order to link the cluster observables to the mass, we made use of the following scaling relations \citep{Ade:2015gva} for the integrated Compton $y$-profile, $Y_{500}$
\begin{equation}\label{eq:Y500}
E^{-\beta} (z) \left[ \dfrac{D_A^2 (z) Y_{500}}{10^{-4} \, \text{Mpc}^2} \right] = Y_* \left[ \dfrac{h}{0.7}\right] ^{-2 + \alpha} \left[ \dfrac{(1-b) M_{500}}{6 \cdot 10^{14} M_{\odot}} \right] ^{\alpha}
\end{equation}
and for the cluster angular size
\begin{equation}\label{eq:theta500}
\theta_{500} = \theta _* \left[ \dfrac{h}{0.7}\right] ^{-2/3} \left[ \dfrac{(1-b) M_{500}}{3 \cdot 10^{14} M_{\odot}} \right]^{1/3} E^{-2/3} (z) \left[ \dfrac{D_A(z)}{500 \, \text{Mpc}}\right] ^{-1} \, .
\end{equation}
In Eqs. \ref{eq:Y500} and \ref{eq:theta500} $D_A(z)$ is the angular diameter distance, $h$ is the reduced Hubble constant $h=H_0/100$, $E(z) = H(z)/H_0$ and $b$ is the mass bias. It is defined through the ratio between the mass estimated assuming hydrostatic equilibrium and the real mass of the cluster, $(1-b) = M_{500,\rm HE}/M_{500}$. Since this is the definition that enters directly into the scaling relations, in the following we report results for the total $(1-b)$ quantity. For the coefficients of the scaling relations we have followed the approach in \cite{Ade:2015gva}. We fixed $\theta_*= 6.997 \, \text{arcmin}$ and $\beta=0.66$, while for $\alpha$, $Y_*$ and the dispersion of the scaling relations, $\sigma _{\ln Y_*}$, we adopted Gaussian priors, as defined in Table 1 of \cite{Ade:2015gva}.

The predicted number counts of galaxy clusters observed in a redshift bin $[z_i,z_{i+1}]$ is given by
\begin{equation}\label{eq:counts1}
n_i = \int _{z_i} ^{z_{i+1}} dz \dfrac{dN}{dz} \, ,
\end{equation}
\noindent where
\begin{equation}\label{eq:counts2}
\dfrac{dN}{dz} = \int d\Omega \int _{M_{\text{min}}} ^{M_{\text{max}}}  dM_{500} \, \hat{\chi} (z,M_{500};l,b)  \dfrac{dN(M_{500},z)}{dM_{500}} \dfrac{dV_c}{dz \, d\Omega} \, .
\end{equation}
In Eq. \ref{eq:counts2}, $\hat{\chi} (z,M_{500};l,b)$ is the survey completeness at the sky position $(l,b)$, $dN(M_{500},z)/dM_{500}$ is the mass function and $dV_c/dz \, d\Omega$ is the volume element. 
The generalization of this description to also define number counts as functions of the signal-to-noise ratio can be found in \cite{Ade:2015fva}.

Assuming the halo model (see e.g. \cite{Cooray:2000ge}), the expected tSZ power spectrum is defined as the sum of one-halo and two-halo terms
\begin{equation}\label{eq:Cell}
C_{\ell}^{\text{tSZ}} = C_{\ell} ^{1\, \text{halo}} + C_{\ell} ^{2\, \text{halo}} \, .
\end{equation}
The one-halo term is expressed as
\begin{eqnarray}\label{eq:Cell1halo}
C_{\ell} ^{1\, \text{halo}} &=& \int _0 ^{z_{\text{max}}} dz \dfrac{dV_c}{dz\, d\Omega} \notag \\
& \times & \int _{M_{\text{min}}} ^{M_{\text{max}}} dM \dfrac{dN(M_{500},z)}{dM_{500}} |\tilde{y}_{\ell}(M_{500},z)| ^2 \notag \\
& \times & \exp{\left( \dfrac{1}{2} \sigma ^2 _{\ln Y^*}  \right)} \, .
\end{eqnarray}
The term $\tilde{y}_{\ell}(M_{500},z)$ in Eq. \ref{eq:Cell1halo} is the Fourier transform on the sphere of the Compton parameter $y$ of individual clusters. It is proportional to the pressure profile, for which we followed the evaluation in \cite{Arnaud:2009tt}.
The two-halo term \citep{Komatsu:1999ev} is defined as
\begin{eqnarray}\label{eq:Cell2halo}
C_{\ell} ^{2\, \text{halo}} &=& \int _0 ^{z_{\text{max}}} dz  \dfrac{dV_c}{dz\, d\Omega}   \notag \\
 & \times & \Bigg[   \int  _{M_{\text{min}}} ^{M_{\text{max}}}  dM \dfrac{dN(M_{500},z)}{dM_{500}} \tilde{y}_{\ell}(M_{500},z)   \notag \\
 & \times &  B(M_{500},z) \Bigg] ^2 \times P(k,z) \,,
\end{eqnarray}
where $P(k,z)$ is the matter power spectrum and $B(M,z)$ is the time-dependent linear bias factor, for which we followed \cite{Komatsu:1999ev}.

We also took into account the contribution from the trispectrum term, in order to correctly evaluate the errors. Following \cite{Cooray:2001wa} and \cite{Komatsu:2002wc}, the dominant term in the halo model is defined as
\begin{eqnarray}\label{eq:T_ll}
T_{\ell \ell'} & \simeq & \int _0 ^{z_{\text{max}}} dz \dfrac{dV_c}{dz d\Omega} \notag \\
& \times & \int _{M_{\text{min}}} ^{M_{\text{max}}} dM \left[ \dfrac{dN(M_{500},z)}{dM_{500}} \right. \notag \\
& \times & |\tilde{y}_{\ell}(M_{500},z)|^2 |\tilde{y}_{\ell'}(M_{500},z)|^2 \biggr] \, .
\end{eqnarray}

\subsection{Mass bias modelling}
We considered a parametric representation of the mass bias, such that the total $(1-b)$ quantity can be defined as
\begin{equation}\label{eq:1mb_par}
(1-b)_{\rm var} (M,z) = (1-\mathcal{B}) \cdot \left( \dfrac{M}{M_*}\right)^{\alpha _b}\cdot \left( \dfrac{1+z}{1+z_*}\right)^{\beta _b} \, ,
\end{equation}
where $(1-\mathcal{B})$ is an amplitude, $M_*=4.82 \cdot 10^{14} M_{\odot}$ is the mean mass value of the Planck cluster sample and $z_*=0.22$ is the median value of the clusters catalogue that we considering. 
We substituted this definition of $(1-b)$ into Eqs. \ref{eq:Y500} and \ref{eq:theta500} for the scaling relations.
Following the analysis in \cite{Salvati:2017rsn}, we applied a WL prior from the Canadian Cluster Comparison Project (CCCP, \citealp{Hoekstra:2015gda}) on the total quantity $(1-b)_{\rm var}$ evaluated at the mean mass and redshift for the sample considered in the CCCP analysis, that is,
 \begin{equation}\label{eq:1mb_par_cccp}
(1-b)_{\rm var} (M_{\rm CCCP},z_{\rm CCCP}) = 0.780 \pm 0.092 \, ,
\end{equation}
with $M_{\rm CCCP} = 14.83 \cdot 10^{14} \, h^{-1} M_{\odot}$ and $z_{\rm CCCP} = 0.246$.

For CMB data, we exploited the new results from the Planck collaboration \citep{Akrami:2018vks,Aghanim:2018eyx}. 
The new Planck likelihood is not yet public. In order to reproduce the constraints on the cosmological parameters $\Omega_m$ (matter density) and $\sigma_8$ (normalization of the matter power spectrum), we mimicked the Planck 2018 likelihood assuming a Gaussian prior on the optical depth, $\tau=0.054 \pm 0.007$ \citep{Aghanim:2018eyx}. We checked that the $68\%$ confidence level (c.l.) constraints that we obtain for $\Omega_m$ and $\sigma_8$ are in agreement with the ones reported in \cite{Aghanim:2018eyx}. 

We stress that tSZ data are not able to constrain the entire set of cosmological parameters alone. For this reason, when not adding CMB data, we considered BAO measurements from \cite{Anderson:2013zyy}.
In the entire analysis we make use of mass function from \cite{Tinker:2008ff}. In particular, we exploited the formulation for the redshift evolution of the mass function, as described in Section 3 of \cite{Tinker:2008ff}, and interpolated in order to obtain the mass function coefficients at $\Delta_c = 500$. 

The results discussed in the next section were obtained using Monte Carlo Markov chains (MCMC). We simultaneously sampled on cosmological and mass bias parameters, together with the other scaling relation parameters, $\alpha$, $Y_*$ and $\sigma _{\ln Y_*}$. We used the November 2016 version of the publicly available package cosmomc \cite{Lewis:2002ah}.
Full details of the way in which we combined the likelihood for tSZ number counts and power spectrum are discussed in \cite{Salvati:2017rsn}. 

In order to check the consistency of our results, we performed several tests. 
We analysed the effect of different mass bias calibrations, applying a Gaussian prior at the mean mass and redshift value for the considered sample (as done for the CCCP calibration). In particular, we considered the WL calibration from the Weighting the Giants project (WtG, \citealp{vonderLinden:2014haa})
 \begin{equation}\label{eq:1mb_par_wtg}
(1-b)_{\rm var} (M_{\rm WtG},z_{\rm WtG}) = 0.688 \pm 0.072 \, ,
\end{equation}
with $M_{\rm WtG} = 13.08 \cdot 10^{14} \, M_{\odot}$ and $ z_{\rm WtG} = 0.31$,
and an evaluation from hydrodynamical simulations, exploiting results from \cite{Biffi:2016yto} (BIFFI)
\begin{equation}
(1-b)_{\rm BIFFI}  (M_{\rm BIFFI},z_{\rm BIFFI})= 0.877 \pm 0.015 \, ,
\end{equation}
with $M_{\rm BIFFI} = 10.53 \cdot 10^{14} \, M_{\odot}$ and $ z_{\rm BIFFI} =0 $.

We then considered a discrete redshift dependence for the mass bias, dividing the entire redshift range of the PSZ2 cosmo sample in different bins, with different mass bias values. We checked how the results change for different redshift and signal-to-noise ratio cuts in the cluster catalogue.

\section{Results}\label{sec:results}

We report the results obtained for the mass-redshift parametrisation of the mass bias. We have analysed only the $\Lambda$CDM scenario of cosmology. Regarding the cosmological parameters, we focussed on the total matter density $\Omega _m$ and the normalization of the matter power spectrum, described through $\sigma _8$. Indeed the other parameters are not affected by this parametrisation.

We define the `standard' scenario as the case in which the mass bias is a varying parameter but without any mass or redshift dependence, i.e. $(1-b)_{\rm var} = (1-\mathcal{B})$, labelled as `$+ \, (1-\mathcal{B})$' only.
We compared this case with the one adding the mass and redshift dependencies, quantified by the $\alpha _b$ and $\beta _b$ parameters. 
When adding the mass-redshift variation, we also considered the case where the amplitude in Eq.~\eqref{eq:1mb_par} is fixed to the value found in the `standard' scenario for the tSZ data combination $C_{\ell}^{\rm tSZ} + \rm NC^{\rm tSZ} + \rm BAO + (1-\mathcal{B})$ , i.e. $(1-\mathcal{B})=0.75$ (see Table~\ref{tab:mz}).
Therefore we fixed  $(1-\mathcal{B})=0.75$ for both $C_{\ell}^{\rm tSZ} + \rm NC^{\rm tSZ} + \rm BAO \, + \,\alpha_b \, + \, \beta_b$ and $C_{\ell}^{\rm tSZ} + \rm NC^{\rm tSZ}  \, + \, \rm CMB \, + \alpha_b \, + \, \beta_b$  scenarios.
We first discuss results for tSZ+BAO data combination and then explore the effect of this mass-redshift parametrisation also on the tSZ+CMB combination.

\subsection{tSZ power spectrum and number counts}

\begin{figure}[!ht]
\centering
\includegraphics[scale=0.225]{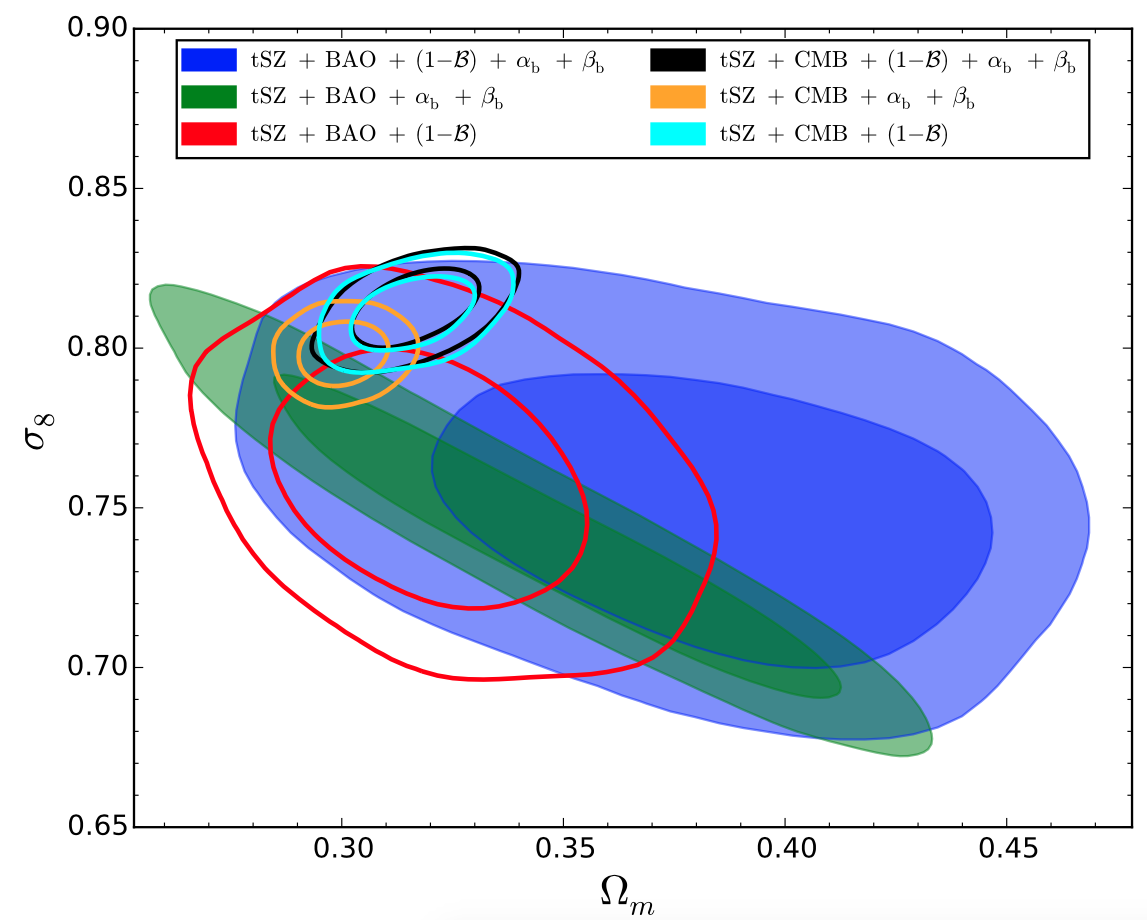}
\caption{\footnotesize{Two-dimensional probability distribution for the parameters $\Omega_m$ and $\sigma_8$, showing $68\%$ and $95\%$ c.l. We show the tSZ+BAO data, with tSZ = $C_{\ell}^{\rm tSZ} + \rm NC^{\rm tSZ}$: varying only the amplitude in Eq.~\eqref{eq:1mb_par} (red contours), adding the mass and redshift dependence (blue filled contours) and fixing the amplitude (green filled contours). We show the same combinations adding CMB data (cyan, black and orange contours).}}
\label{fig:mz_omegam_sigma8}
\end{figure}

We discuss results for the $C_{\ell}^{\rm tSZ} + \rm NC^{\rm tSZ} + \rm BAO$ data combination. The two-dimensional probability distributions in the $(\Omega_m,\sigma_8)$ parameter plane are shown in Fig.~\ref{fig:mz_omegam_sigma8} for the standard scenario (`$+ \, (1-\mathcal{B})$', red contours), allowing for the mass and redshift variation (`$+\alpha_b + \beta_b$', green filled contours) and for the complete parameters combination (`$+ \, (1-\mathcal{B})+\alpha_b + \beta_b$', blue filled contours). The $68\%$ c.l. constraints are reported in Table~\ref{tab:mz} for all the parameters combinations.

\begin{figure}[!ht]
\centering
\includegraphics[scale=0.225]{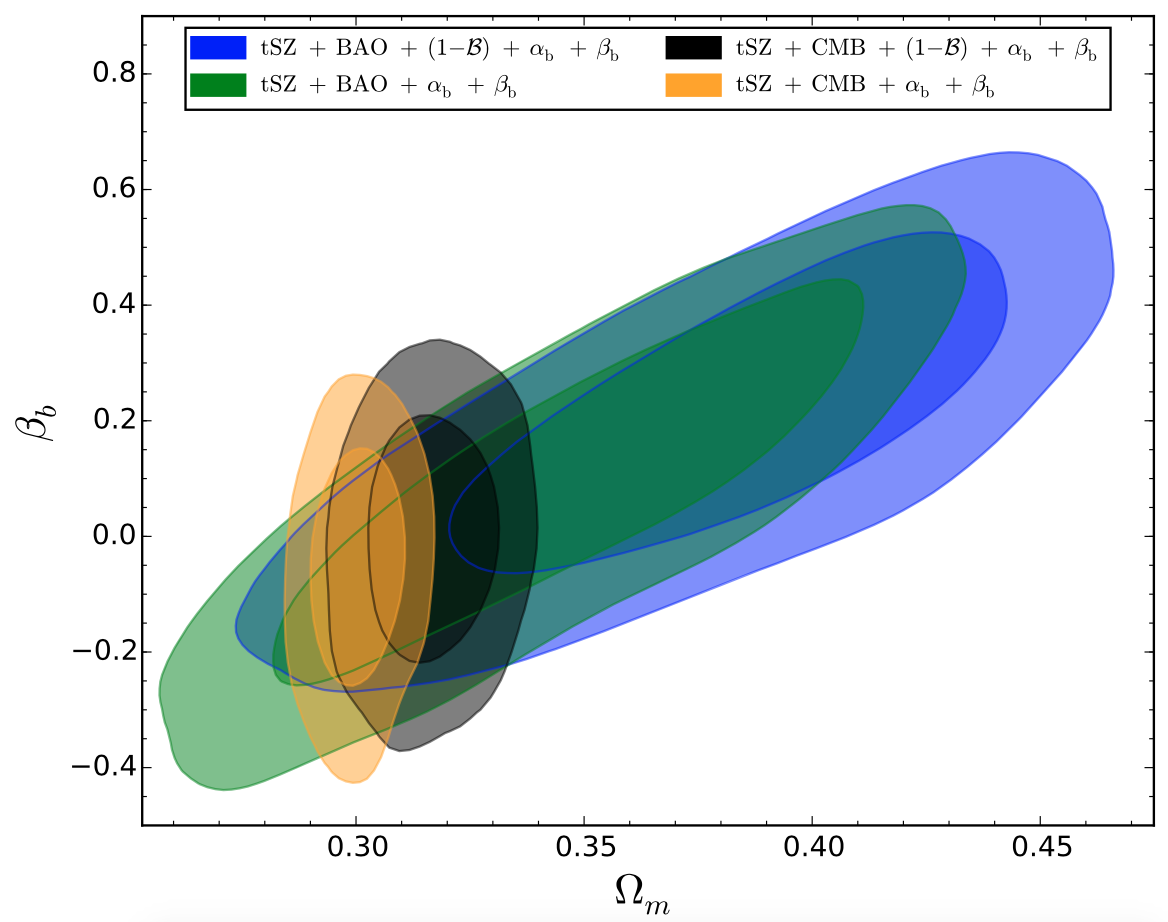}
\caption{\footnotesize{Two-dimensional probability distributions for the parameters $\Omega_m$ and $\beta_b$, showing $68\%$ and $95\%$ c.l. We show tSZ+BAO results, with tSZ = $C_{\ell}^{\rm tSZ} + \rm NC^{\rm tSZ}$, for the complete mass bias parameters combination (blue) and fixing the amplitude (green). We report the same combinations when also adding CMB data (black and orange respectively).}}
\label{fig:mz_omegam_betab}
\end{figure}

We began by comparing the results for the standard scenario with those obtained when varying all mass bias parameters. In the latter case, we see an enlargement of the constraints towards higher values of $\Omega_m$. This is mainly due to the degeneracy between the matter density and the redshift dependence, as shown in Fig.~\ref{fig:mz_omegam_betab}. Regarding the mass and redshift dependence, we first focussed on the constraints on $\alpha_b$ and $\beta_b$. For the mass variation, we obtain $\alpha_b = 0.141 _{-0.092}^{+0.083}$. We stress that the $\alpha_b$ parameter is strongly correlated with the amplitude of the parametrisation, $(1-\mathcal{B})$, as it can be seen in the scatter plot in Fig.~\ref{fig:mz_alpha_beta_scatter}, left panel. Therefore we are not able to fully disentangle the contributions of the two parameters to evaluate the exact mass dependence. For the redshift variation, we find $\beta_b = 0.20 _{-0.16}^{+0.20}$. 

We quantify the effect of these results on the total varying mass bias, $(1-b)_{\rm var}$. We report in Fig.~\ref{fig:mz_bvar_Mz} (green and blue curves) the trend of $(1-b)_{\rm var}$ with respect to redshift and mass, where the shaded areas represent the estimated $1 \, \sigma$ error. In order to stress the effect of the mass and redshift variation, we show the trend for $(1-b)_{\rm var}$ at fixed values of redshift and mass respectively, considering in each case the lowest and highest values in the cluster catalogue range. From these plots we can see that the total mass bias shows a mild variation in mass and redshift. Therefore, even if the $\alpha_b$ and $\beta_b$ constraints are marginally consistent (within $2 \, \sigma$) with 0, we still see a hint for mass and redshift evolution. In particular, we stress that $(1-b)_{\rm var}$ increases with redshift, therefore providing a decreasing trend for the mass bias.
In order to further understand this particular trend, we performed several consistency checks. We report the results in the Section \ref{sec:tests} below.

We conclude by reporting the results obtained when fixing the amplitude $(1-\mathcal{B}) = 0.75$. We chose this value for the amplitude since it is the preferred one from the standard scenario (`$+(1-\mathcal{B})$'). Allowing for $\alpha_b$ and $\beta_b$ to vary enlarges the constraints in the ($\Omega_m,\sigma_8$) parameter plane, as for the complete combination `$(1-\mathcal{B}) + \alpha_b + \beta_b$'. However, the fixed amplitude moves the constraints of the mass-redshift parametrisation back to being completely consistent with 0. Indeed, we find $\alpha_b = 0.052 \pm 0.046 $ and $\beta_b = 0.09 _{-0.18}^{+0.21} $.

\subsection{Comparison between CMB and tSZ data}
We start comparing results on the $(\Omega _m, \sigma _8)$ parameter plane for the `standard' (`$+ \, (1-\mathcal{B})$', only) scenario. \cite{Aghanim:2018eyx} provides the following $68\%$ c.l. constraints, from the baseline \texttt{Planck} TT,TE,EE+lowE+lensing: $\Omega _m= 0.315 \pm 0.007$ and $\sigma _8=0.811 \pm 0.006$. Regarding the combination with tSZ data (considering number counts alone in this case), they provide the value for the mass bias $(1-\mathcal{B})=0.62 \pm 0.03$. Comparing these values with those reported here in Table~\ref{tab:mz} for the $C_{\ell}^{\rm tSZ} + \rm NC^{\rm tSZ} + \rm CMB \, + \, (1-\mathcal{B})$ scenario, we find a complete agreement, showing the significantly stronger constraining power of CMB with respect to tSZ probes.

Two-dimensional probability distributions in the $(\Omega_m, \sigma_8)$ parameter plane are shown in Fig.~\ref{fig:mz_omegam_sigma8} for the standard scenario (`$+ \, (1-\mathcal{B})$', cyan contours), opening for the mass and redshift variation (`$+\alpha_b + \beta_b$', orange contours) and for the complete parameters combination (`$+ \, (1-\mathcal{B})+\alpha_b + \beta_b$', black contours). The $68\%$ c.l. constraints are reported in Table~\ref{tab:mz}.

Next we compare results for $C_{\ell}^{\rm tSZ} + \rm NC^{\rm tSZ} + \rm CMB \, + \, (1-\mathcal{B})$ with those considering tSZ+BAO data, i.e. $C_{\ell}^{\rm tSZ} + \rm NC^{\rm tSZ} + \rm BAO \, + \, (1-\mathcal{B})$. As discussed in \cite{Salvati:2017rsn} and \cite{Aghanim:2018eyx}, the lower value of the optical depth $\tau$ helps in reducing the discrepancy between tSZ and CMB data, at least on the $\sigma _8$ parameter. Given the degeneracy between the $\sigma_8$ parameter and the mass bias, the lower value of $\tau$ helps also in partly reducing the discrepancy on this latter quantity. 
Nevertheless, current tSZ+CMB data combination prefers a value of the mass bias that is still only marginally consistent with current simulations and WL calibrations, see for example, the collection of results reported in \citet[Fig.~10]{Salvati:2017rsn}. Indeed, we find $(1-\mathcal{B})=0.62 \pm 0.04$.

We therefore analysed whether the mass-redshift parametrisation proposed in Eq.~\ref{eq:1mb_par} can allow for a convergence of CMB and tSZ results towards higher values of the mass bias. The results from $C_{\ell}^{\rm tSZ} + \rm NC^{\rm tSZ} + \rm CMB$ data combination confirm the strongest constraining power of CMB data. Indeed, for the complete case with varying $(1-\mathcal{B}) \, + \alpha_b \, + \, \beta_b$, results are completely in agreement with the standard scenario where only $(1-\mathcal{B})$ is varying, with $\alpha_b$ and $\beta_b$ being consistent with $0$. 
The two dataset combinations also provide the same $\chi^2$ value ($\chi^2_{(1-\mathcal{B})} = \chi^2 _{(1-\mathcal{B}) + \alpha + \beta} = 2956$)

We then explored the results when fixing the amplitude in Eq.~\eqref{eq:1mb_par} to $(1-\mathcal{B})=0.75$. 
Our aim is to verify whether the preferred low constraints for the mass bias from CMB could be due to the assumption of a constant value in the entire mass and redshift range, in other words, if a possible mass-redshift variation can allow for a mass bias around 0.75 at some point in the mass-redshift space.
The `$+\alpha+\beta$' parameter combination produces a shift in the constraints in the $(\Omega_m,\sigma_8)$ plane, with $\chi^2 _{\alpha + \beta} = 2962$. We therefore compared this model with the standard `$+(1-\mathcal{B})$' one, applying the Bayesian Inference Criterion (BIC, \cite{Schwarz:1978tpv}). The $\Delta$ BIC between the `$+\alpha+\beta$' and $+(1-\mathcal{B})$ models is $\Delta \rm BIC \simeq 12.5$, therefore providing strong evidence for the first model to be disfavoured.

\begin{figure*}[!h]
  \centering
  \subfigure{\includegraphics[scale=0.46]{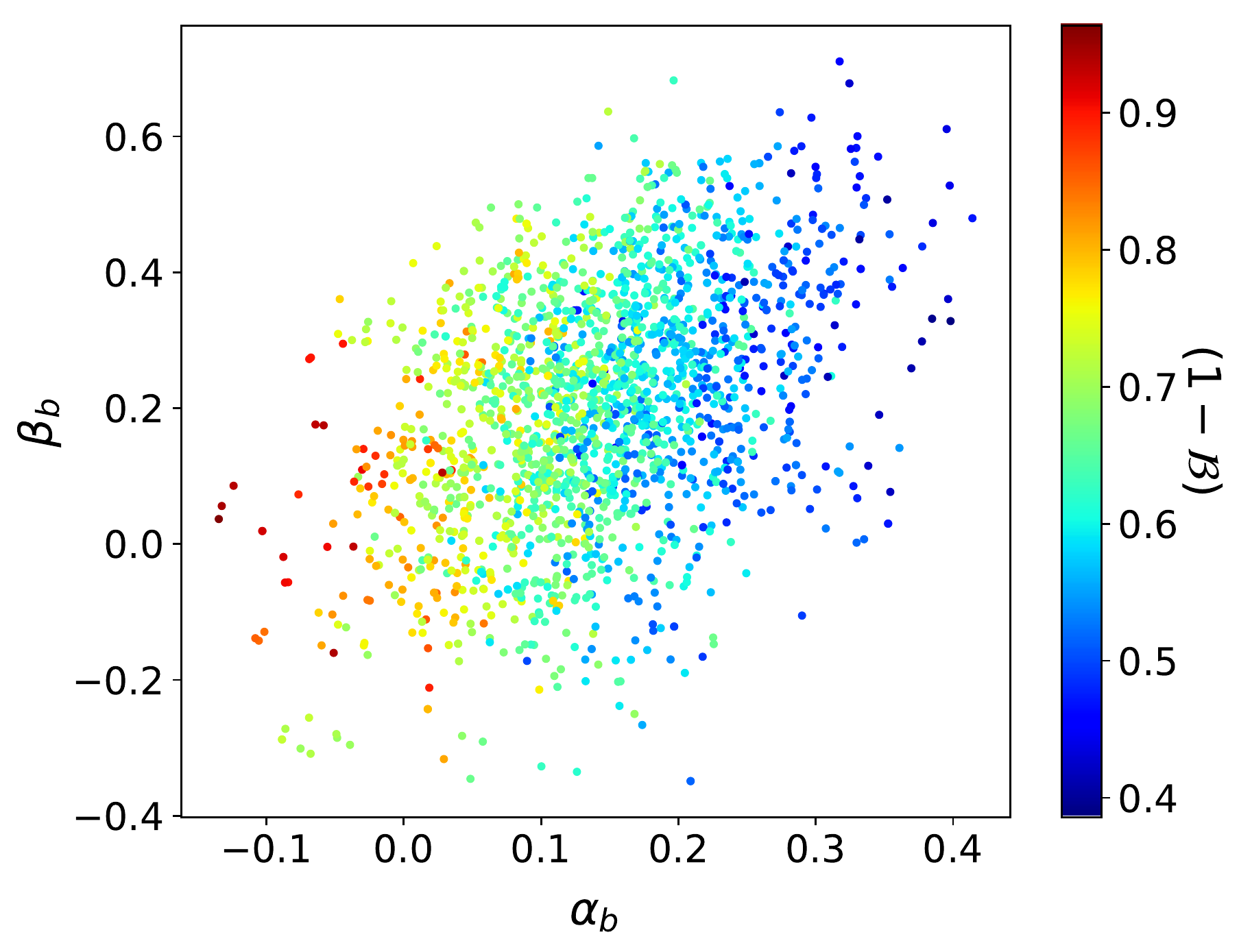}} \,
  \subfigure{\includegraphics[scale=0.46]{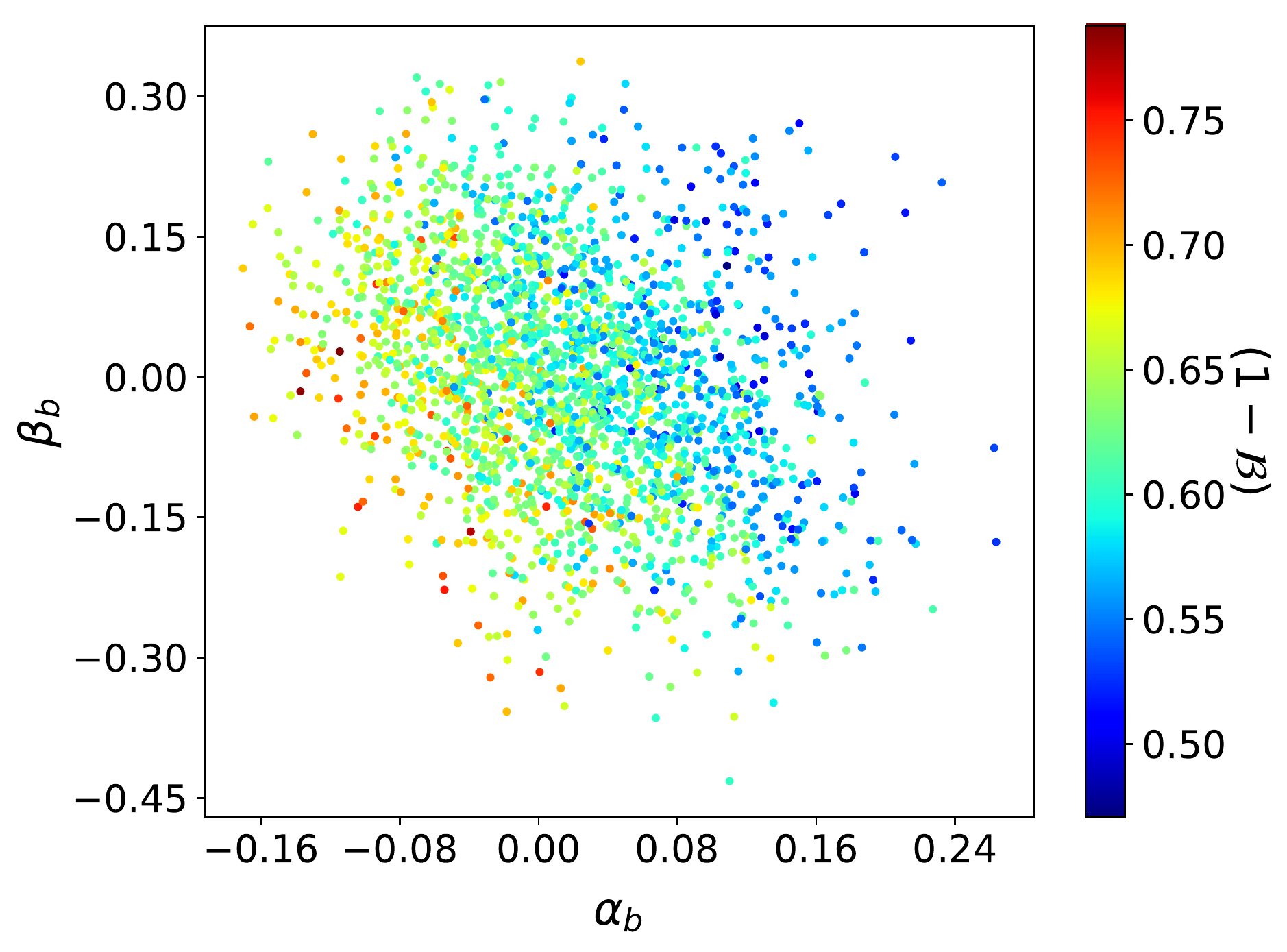}}
  \caption{\footnotesize{ $(\alpha_b,\beta_b)$ parameters at different values of the amplitude $(1-\mathcal{B})$. We show results for $C_{\ell}^{\rm tSZ} + \rm NC^{\rm tSZ} + \rm BAO \, + \, (1-\mathcal{B}) \, + \alpha_b \, + \, \beta_b$ (left panel) and $C_{\ell}^{\rm tSZ} + \rm NC^{\rm tSZ} + \rm CMB \, + \,(1-\mathcal{B}) \, + \alpha_b \, + \, \beta_b$ (right panel).}}
  \label{fig:mz_alpha_beta_scatter}
\end{figure*}

In the right panel of Fig.~\ref{fig:mz_alpha_beta_scatter}, we report the two-dimensional probability distribution for $(\alpha_b,\beta_b)$ at different values of the amplitude $(1-\mathcal{B})$. We find a mild correlation between $\alpha_b$ and $\beta_b$. Furthermore, we see that we are never able to reach values of $(1-\mathcal{B}) \sim 0.8$, for any combination of $(\alpha_b,\beta_b)$.

\begin{figure*}[!h]
  \centering
  \subfigure{\includegraphics[scale=0.446]{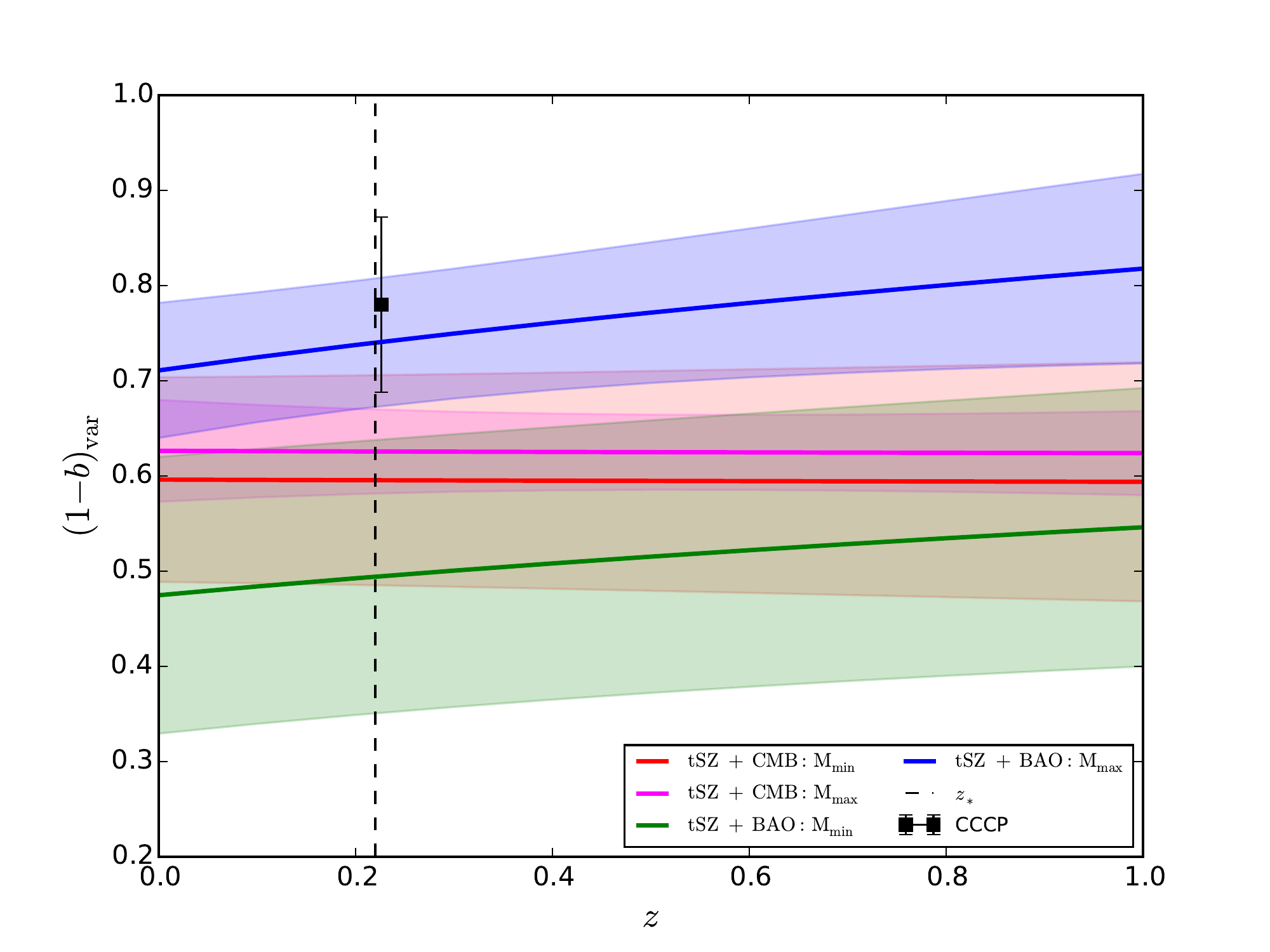}} \,
  \subfigure{\includegraphics[scale=0.446]{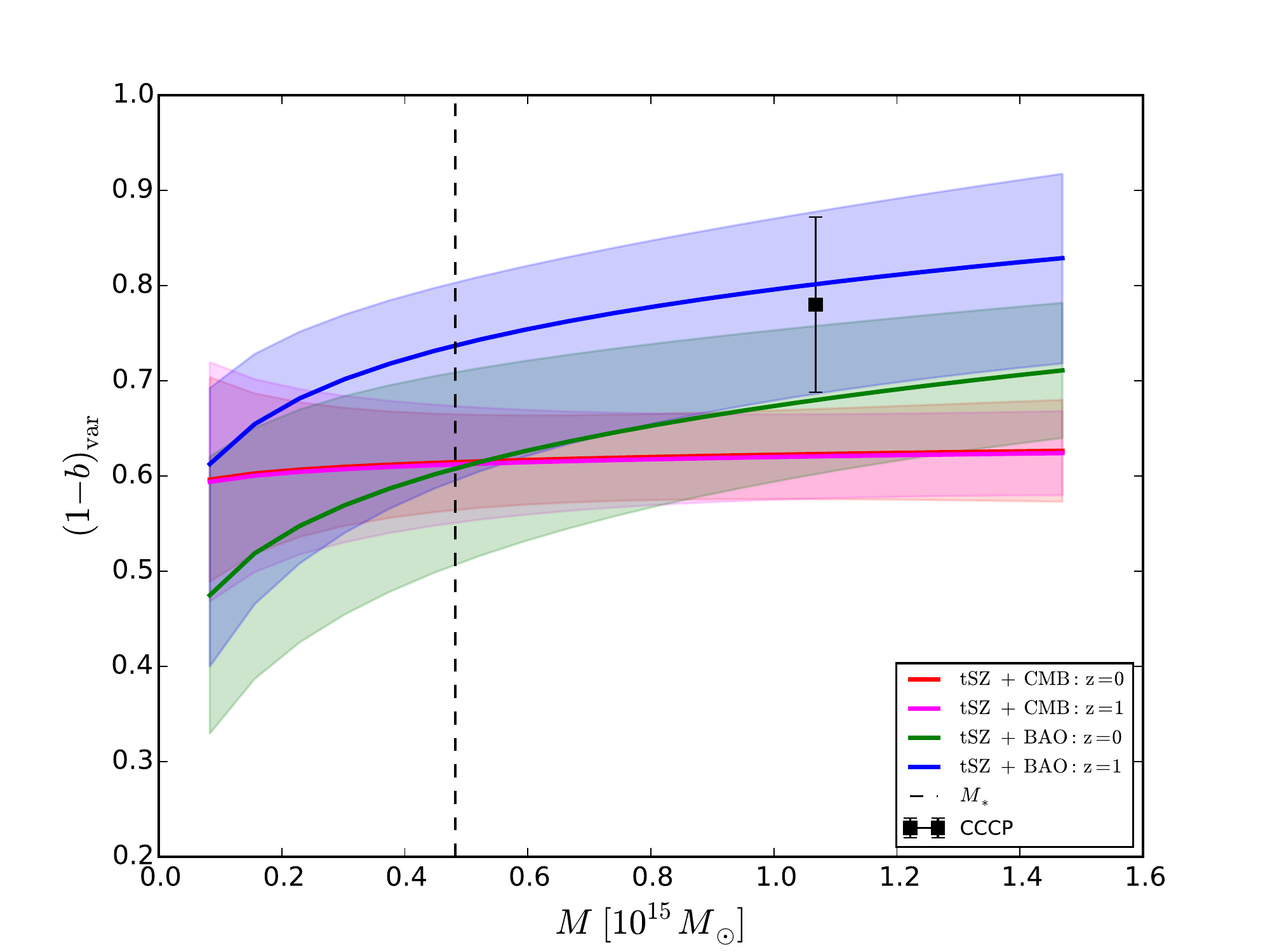}}
  \caption{\footnotesize{Redshift (left panel) and mass (right panel) variation of $(1-b)_{\rm var}$ at fixed values of mass and redshift respectively. We show results for the combinations $C_{\ell}^{\rm tSZ} + \rm NC^{\rm tSZ} + \rm BAO \, + \, (1-\mathcal{B}) \, + \alpha_b \, + \, \beta_b$ (labelled `tSZ + BAO', green and blue) and $C_{\ell}^{\rm tSZ} + \rm NC^{\rm tSZ} + \rm CMB\, + \, (1-\mathcal{B}) \, + \alpha_b \, + \, \beta_b$ (labelled as `tSZ+CMB', red and magenta).}}
  \label{fig:mz_bvar_Mz}
\end{figure*}

We conclude by discussing the results for the total $(1-b)_{\rm var}$ quantity, for the complete $(1-\mathcal{B}) \, + \alpha_b \, + \, \beta_b$ scenario. In Fig.~\ref{fig:mz_bvar_Mz} we show the $(1-b)_{\rm var}$ trend with respect to redshift and mass, for tSZ+CMB data combination (red and magenta lines and shaded areas). As described in previous section, we show the trend for $(1-b)_{\rm var}$ at fixed values of mass and redshift respectively, considering in each case the lowest and highest values in the cluster catalogue range.
From these results we can see that adding CMB data keeps the value of $(1-b)_{\rm var}$ almost constant in mass and redshift, while the results for tSZ data show a variation. Furthermore, it is clear that the adopted mass-redshift parametrisation does not help in solving the remaining tension on the mass bias between CMB and tSZ data.

$ $ \\

\begin{table*}
\begin{center}
\scalebox{0.93}{
\begin{tabular}{c|c|c|c|c|c}
\hline
\hline
\morehorsp
 Datasets & $\Omega _m$ & $\sigma _8$ & $(1-\mathcal{B})$ & $\alpha _b $ & $\beta _b$\\
\hline
\morehorsp
 $C_{\ell}^{\rm tSZ} + \rm NC^{\rm tSZ} + \rm BAO \, + \, (1-\mathcal{B})$ &$ 0.322 _{-0.025}^{+0.018} $ &$ 0.759 \pm 0.024$ &$0.753 \pm 0.064$ &$0 $ &$ 0$ \\
 \hline
\morehorsp
$C_{\ell}^{\rm tSZ} + \rm NC^{\rm tSZ} + \rm BAO \, + \, (1-\mathcal{B}) \, + \alpha_b \, + \, \beta_b$ &$0.380 _{-0.032}^{+0.048} $ &$0.751 _{-0.032}^{+0.024} $ &$0.63 \pm 0.10 $ &$0.141 _{-0.092}^{+0.083} $ &$ 0.20 _{-0.16}^{+0.20}$ \\
 \hline
\morehorsp
$C_{\ell}^{\rm tSZ} + \rm NC^{\rm tSZ} + \rm BAO \, + \,\alpha_b \, + \, \beta_b$ &$0.345 \pm 0.040 $ &$0.742 _{-0.035}^{+0.026} $ &$0.75 $ &$0.052 \pm 0.046 $ &$0.09 _{-0.18}^{+0.21} $ \\
\hline
\hline
\morehorsp
$C_{\ell}^{\rm tSZ} + \rm NC^{\rm tSZ} \, + \, \rm CMB \, + \, (1-\mathcal{B})$ &$0.316 \pm 0.009 $ &$0.811 \pm 0.007 $ &$0.622 \pm 0.033 $ &$ 0$ &$ 0$ \\
\hline
\morehorsp
$C_{\ell}^{\rm tSZ} + \rm NC^{\rm tSZ} \, + \, \rm CMB \, + \, (1-\mathcal{B}) \, + \alpha_b \, + \, \beta_b$ &$0.316 \pm 0.009 $ &$0.812 \pm 0.008 $ &$0.614 _{-0.052}^{+0.045} $ &$0.017_{-0.085}^{+0.064} $ &$0 _{-0.13}^{+0.14} $ \\
\hline
\morehorsp
$C_{\ell}^{\rm tSZ} + \rm NC^{\rm tSZ}  \, + \, \rm CMB \, + \alpha_b \, + \, \beta_b$ &$0.300 \pm 0.006 $ &$ 0.799 \pm 0.006$ &$0.75 $ &$-0.066 _{-0.063}^{+0.051} $ &$-0.06 _{-0.12}^{+0.13} $ \\
\hline
\end{tabular}}
\caption{\footnotesize{$68\%$ c.l. constraints for cosmological and mass bias parameters, for the different dataset combinations.}}
\label{tab:mz}
\end{center}
\end{table*}

\section{Robustness tests}\label{sec:tests}
To further understand the hint of redshift dependence leading to increasing values of $(1-b)_{\rm var}$ when considering the $C_{\ell}^{\rm tSZ} + \rm NC^{\rm tSZ} + \rm BAO \, + \, (1-\mathcal{B}) \, + \alpha_b \, + \, \beta_b$ dataset combination, we perform several analyses. Our aim is to understand if the obtained constraints may depend on the choice of the mass bias calibration, the mass-redshift dependence description or the sample selection.

Before showing these results, we analysed the effect of the tSZ power spectrum. Indeed, the wider redshift range covered by tSZ power spectrum measurements ($z=[0,3]$) could be the cause for the increasing trend of $(1-b)_{\rm var}$ with redshift.
We compared the results for the complete $C_{\ell}^{\rm tSZ} + \rm NC^{\rm tSZ} + \rm BAO \, + \, (1-\mathcal{B}) \, + \alpha_b \, + \, \beta_b$ combination with those obtained from $\rm NC^{\rm tSZ} + \rm BAO \, + \, (1-\mathcal{B}) \, + \alpha_b \, + \, \beta_b$, the latter being reported in Table~\ref{tab:mz_tests}. From this comparison it is clear that results are driven by the stronger constraining power of tSZ number counts. Indeed, the addition of tSZ power spectrum only helps in slightly improving the constraints precision, not changing the general results.
Therefore we confirm that, in the `$(1-\mathcal{B}) \, + \alpha_b \, + \, \beta_b$' scenario too, tSZ power spectrum provides a lower constraining power with respect to number counts, as it is already shown for the `standard' scenario in \cite{Salvati:2017rsn}.

Hence, for all the following tests we decided to focus on the $\rm NC^{\rm tSZ} + \rm BAO$ combination alone, considering as baseline results obtained with CCCP prior on $(1-b)_{\rm var}$.
As a reference, in Table~\ref{tab:mz_tests} we also report the results for the `standard' scenario (i.e. fixing $\alpha_b = 0$ and $\beta_b = 0$) for $\rm NC^{\rm tSZ} + \rm BAO$.

\subsection{Effect of external calibrations on $(1-b)_{\rm var}$}

We compared results of the mass-redshift parametrisation for different mass bias calibrations. Results are reported in Table~\ref{tab:mz_tests}, upper panel. We began by considering a significantly different WL calibration, exploiting results from the Weighting the Giants (WtG) analysis \citep{vonderLinden:2014haa}. We applied the calibration at the mean mass and redshift for the WtG catalog, that is 
\begin{equation}
(1-b)_{\rm WtG}  (M_{\rm WtG},z_{\rm WtG})= 0.688 \pm 0.072 \, .
\end{equation}
We chose this calibration, since it provides a higher value of the mass bias (i.e. lower value for $(1-b)$), more in agreement with the expected value from the $C_{\ell}^{\rm tSZ} + \rm NC^{\rm tSZ} + \rm CMB$ combination. We find a general agreement with results obtained when using CCCP calibration. In particular, we find the same hint of redshift dependence, having $\beta_b = 0.26 _{-0.17}^{+0.23}$ for WtG.

We then considered an estimation of the mass bias from cosmological hydrodynamical simulations. We considered the analysis of \cite{Biffi:2016yto}, which gives the value
\begin{equation}
(1-b)_{\rm BIFFI}  (M_{\rm BIFFI},z_{\rm BIFFI})= 0.877 \pm 0.015
\end{equation}
at redshift $z=0$. The lower value of the mass bias is pointing towards lower value of $\sigma_8$, as expected, but still providing consisting constraints for the redshift dependence, with $\beta_b = 0.27 _{-0.17}^{+0.22}$.

We stress how these calibrations are based on different approaches and are evaluated at different redshift and mass (at which we apply the prior in our analysis), having $z_{\rm CCCP} = 0.246$, $z_{\rm WtG} = 0.31$ and $z_{\rm BIFFI} = 0$, $M_{\rm CCCP} = 14.83 \cdot 10^{14} \, h^{-1} M_{\odot}$, $M_{\rm WtG} = 13.08 \cdot 10^{14} M_{\odot}$ and $M_{\rm BIFFI} = 10.53 \cdot 10^{14} M_{\odot}$.
Nevertheless, they provide the same hint of redshift evolution, with the total $(1-b)_{\rm var}$ increasing for higher redshift values. The shape of the mass bias $(1-b)_{\rm var}$ is almost independent on the mass-redshift pivots for the different priors. Indeed, the calibrations simply change the constraints on $\Omega_m$, $\sigma_8$ and $(1-\mathcal{B})$, as reported in Table~\ref{tab:mz_tests}. In Appendix \ref{sec:app} we report a triangular plot, showing this behaviour for the cosmological and mass bias parameters for the different calibrations (Fig.~\ref{fig:biasCAL}).

As a further comparison, we analysed the results when considering only a flat prior for the mass bias, with $(1-b)_{\rm var}=[0.6,1.0]$. Forcing the total mass bias within these limits largely reduces the allowed range for the mass-redshift variation. As shown, for example in Fig.~\ref{fig:mz_bvar_Mz}, the values of $(1-b)_{\rm var}$ for $M_{\rm min}$ and $z=0$ are well below $0.6$. Therefore, we obtain constraints on $\alpha_b$ and $\beta_b$ completely consistent with $0$, as reported in Table~\ref{tab:mz_tests}. Nevertheless, we highlight that even without adding external WL/hydrodynamical calibrations, we find consistent high value for the mass bias, having $(1-\mathcal{B}) = 0.756_{-0.083}^{+0.056}$.

We checked that we obtain consistent results when considering the complete $C_{\ell}^{\rm tSZ} + \rm NC^{\rm tSZ} + \rm BAO$ combination for all the different mass bias calibrations and when applying the flat prior on $(1-b)_{\rm var}$.

\subsection{Binning in redshift}

Next we analysed results from a discrete redshift dependence for the mass bias, dividing the redshift range, $z=[0,1]$, in three bins. In this case too, 
we considered the $\rm NC^{\rm tSZ} + \rm BAO$ combination with CCCP calibration as the baseline.
For this analysis, in each redshift bin, we did not re-evaluate the CCCP cluster masses, but we find the value of $(1-b)$ that allows for consistency between WL and tSZ results.
We chose the binning considering the redshift distributions of the CCCP sub-sample, such that the majority of the clusters is included in the second bin, that also encompasses the mean redshift of the sample. We then mimicked the calibration by adding a gaussian prior on the mass bias parameter in the second redshift bin. The number of clusters in the chosen redshift bins are reported in Table~\ref{tab:zbins}, with the relative prior for the mass bias.
Results are reported in the lower panel of Table~\ref{tab:mz_tests}.

\begin{table}[!h]
\begin{center}
\scalebox{0.8}{
\begin{tabular}{c|c|c|c|c}
\hline
\hline
\morehorsp
  & bin 1 & bin 2 & bin 3 & \multirow{2}*{$(1-b)_2$} \\
\morehorsp
 & $[0,0.2]$ & $[0.2,0.5]$ & $[0.5,1]$ & \\
\hline
\morehorsp
 CCCP &$ 6$ &$11 $ &$1 $ &$0.78 \pm 0.092$\\
\hline
\hline
\morehorsp
PSZ2 cosmo &$ 209$ &$ 200$ &$ 23$ & \\ 
\morehorsp
$M \, [ 10^{14} M_{\odot}]$& $[0.83,8.86]$ & $[3.70,14.69]$ & $[5.87,11.49]$ \\
\hline
\end{tabular}}
\caption{\footnotesize{Number of galaxy clusters per redshift bin, for the different WL calibrations, compared with the Planck cosmological sample \citep{Ade:2015fva}, labelled as `PSZ2 cosmo'. We also report the range of cluster mass for each redshift bin.}}
\label{tab:zbins}
\end{center}
\end{table}

\begin{figure}
\centering
\includegraphics[scale=0.315]{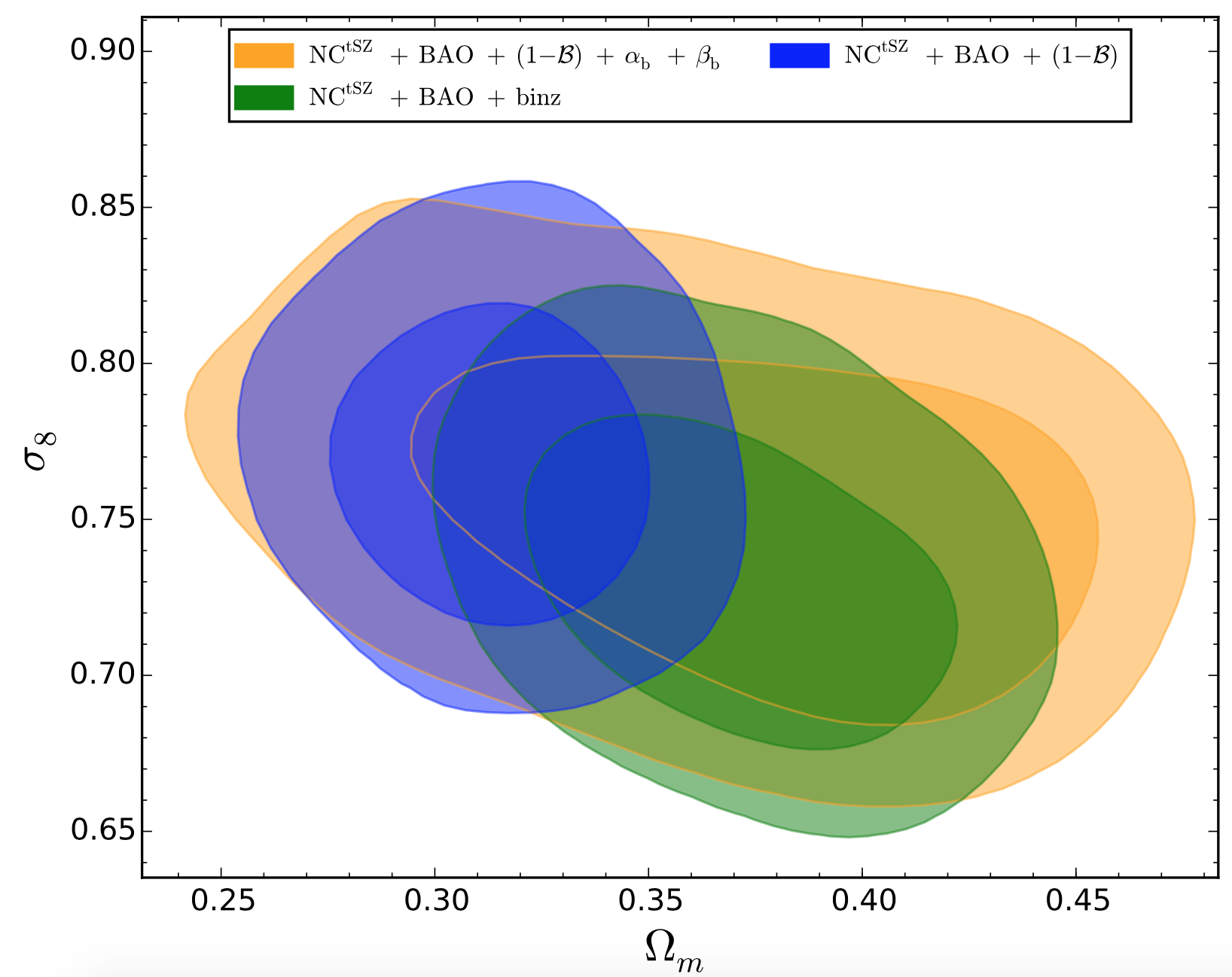}
\caption{\footnotesize{Two-dimensional probability distributions for $\Omega_m$ and $\sigma_8$. We show the comparison between the complete combination $\rm NC^{\rm tSZ} + \rm BAO \, + \, (1-\mathcal{B}) \, + \alpha_b \, + \, \beta_b$ (orange) and results from the binning in redshift (green). As a reference, we show in blue the constraints for the `standard' case $\rm NC^{\rm tSZ} + \rm BAO \, + \, (1-\mathcal{B})$. All results are obtained considering the CCCP calibration.}}
\label{fig:CCCP_omegam_sigma8}
\end{figure}

\begin{figure}
\centering
\includegraphics[scale=0.42]{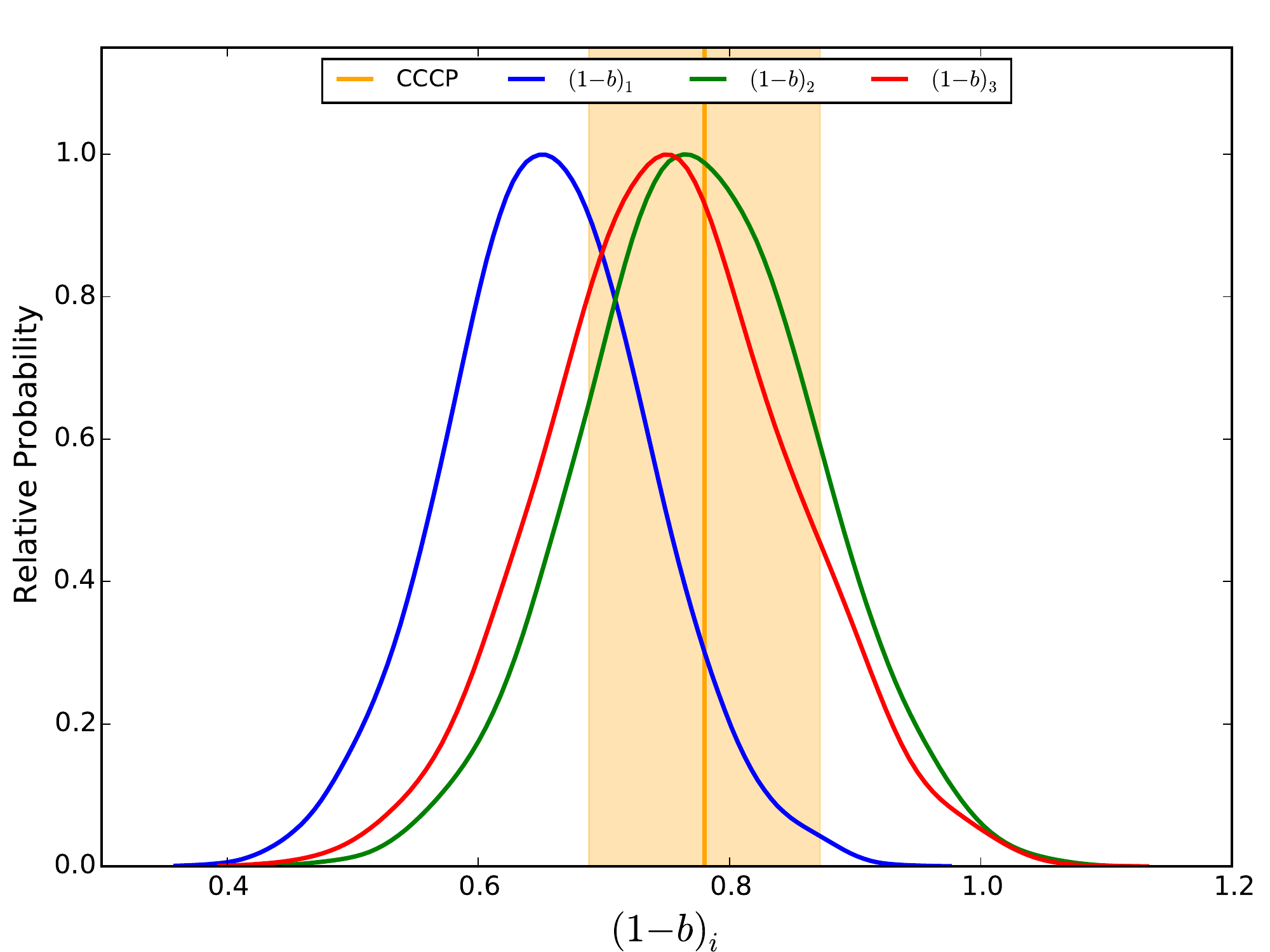}
\caption{\footnotesize{One-dimensional posterior distributions for the mass bias parameter in each redshift bin, for the $\rm NC^{\rm tSZ} + \rm BAO$ combination. As a reference, we show the $1 \, \sigma$ limits for the CCCP calibrations in orange (shaded area).}}
\label{fig:binz_bias}
\end{figure}

We compared this redshift binning analysis with the mass-redshift parametrisation when using the CCCP calibration. In Fig.~\ref{fig:CCCP_omegam_sigma8} we show the two dimensional probability distributions in the $(\Omega_m,\sigma_8)$ plane for the two cases. We find that also the redshift binning analysis provide constraints for $\Omega_m$ shifted towards higher values. We next focussed on the constraints for the mass bias parameters in each redshift bin. The one-dimensional probability distributions are shown in Fig.~\ref{fig:binz_bias}.
We find a mild confirmation for the increasing trend of $(1-b)$ for higher redshift. Indeed, results for the second and third bin point toward higher values, showing only a marginal consistency, within $2 \, \sigma$, with the first bin.
We checked that we obtain consistent results on the shift of the $(1-b)_i$ parameters when considering the complete $C_{\ell}^{\rm tSZ} + \rm NC^{\rm tSZ} + \rm BAO$.

\begin{table*}
\begin{center}
\scalebox{0.9}{
\begin{tabular}{c|c|c|c|c|c}
\hline
\hline
\morehorsp
 Datasets & $\Omega _m$ & $\sigma _8$ &  $(1-\mathcal{B})$ & $\alpha _b $ & $\beta _b$\\
 \hline
\morehorsp
CCCP: $\rm NC^{\rm tSZ} + \rm BAO \, + \, (1-\mathcal{B})$ &$ 0.314 \pm 0.023$ &$0.769 _{-0.034}^{+0.028} $ &$0.751 \pm 0.093 $ &$ 0$ &$0$ \\
\hline
\hline
\morehorsp
CCCP: $\rm NC^{\rm tSZ} + \rm BAO \, + \, (1-\mathcal{B}) \, + \alpha_b \, + \, \beta_b$ &$0.374 _{-0.035}^{+0.065} $ &$0.752 _{-0.040}^{+0.033} $ &$ 0.66_{-0.14}^{+0.10}$ &$ 0.10 \pm 0.10$ &$0.24_{-0.18}^{+0.24} $ \\
\hline
\morehorsp
WtG: $\rm NC^{\rm tSZ} + \rm BAO \, + \, (1-\mathcal{B}) \, + \alpha_b \, + \, \beta_b$ &$ 0.385 _{-0.032}^{+0.060}$ &$0.765 _{-0.035}^{+0.026} $ &$0.597_{-0.094}^{+0.077} $ &$0.11 \pm 0.10 $ &$0.26 _{-0.17}^{+0.23} $  \\
\hline
\morehorsp
BIFFI: $\rm NC^{\rm tSZ} + \rm BAO \, + \, (1-\mathcal{B}) \, + \alpha_b \, + \, \beta_b$ &$0.362 _{-0.023}^{+0.044}$ &$0.699 _{-0.033}^{+0.020} $ &$0.845 \pm 0.063 $ &$ 0.11 _{-0.10}^{+0.08}$ &$ 0.27 _{-0.17}^{+0.22}$ \\
\hline
\morehorsp
flat [0.6,1.0]: $\rm NC^{\rm tSZ} + \rm BAO \, + \, (1-\mathcal{B}) \, + \alpha_b \, + \, \beta_b$ &$0.330 \pm 0.038 $ &$0.753 _{-0.031}^{+0.026} $ &$0.756_{-0.083}^{+0.056} $ &$0.005 _{-0.026}^{+0.029} $ &$ 0.10 \pm 0.16$ \\
\hline
\hline
\morehorsp
 & $\Omega _m$ & $\sigma _8$ &  $(1-b)_1$ & $(1-b)_2 $ & $(1-b)_3$\\
\hline
\morehorsp
CCCP: $\rm NC^{\rm tSZ} + \rm BAO $ &$0.371 _{-0.034}^{+0.028} $ &$0.733_{-0.037}^{+0.028} $ &$0.655 \pm 0.078 $ &$0.775 \pm 0.092 $ &$0.751 \pm 0.095 $ \\
\hline
\end{tabular}}
\caption{\footnotesize{$68\%$ c.l. constraints for cosmological and mass bias parameters, for the different dataset combinations.}}
\label{tab:mz_tests}
\end{center}
\end{table*}

\subsection{Selection effects}

We conclude this section by taking into account possible catalogue selection effects. In particular, we have considered different sub-samples of the entire PSZ2 cosmo sample. As described in Section \ref{sec:method}, the PSZ2 cosmo sample is selected with a signal-to-noise threshold $q_{\rm min} =6$ and counts 439 clusters. As a comparison, following the analysis in \cite{Ade:2015fva}, we considered two other thresholds, $q_{\rm min}=7$ and  $q_{\rm min}=8.5$, which provide samples of 339 and 216 clusters respectively. We label these samples `PSZ2 A' and `PSZ2 B'. We also considered a different range in redshift, selecting only clusters with $z \gtrsim 0.2$. This cut provides a sample of 225 clusters, which we label `PSZ2 C'. As a reference, the specifics of the different samples are reported in Table~\ref{tab:cat}.

\begin{table}[!h]
\begin{center}
\scalebox{0.8}{
\begin{tabular}{c|c|c|c|c}
\hline
\hline
\morehorsp
Catalogue  & $q_{\rm min} $ & $z$ range & $M \, [10^{14}M_{\odot}]$ & n$^{\circ}$ of clusters \\
\hline
\morehorsp
 PSZ2 cosmo &$ 6$ &$[0,1] $ & $[0.83,14.69]$ &$439 $ \\
\hline
\morehorsp
PSZ2 A &$ 7$ &$ [0,1]$ & $ [0.83,14.69]$&$ 339$ \\
\hline
\morehorsp
PSZ2 B &$ 8.5$ &$ [0,1]$ & $ [1.32,14.69]$&$ 216$  \\ 
\hline
\morehorsp
PSZ2 C &$ 6 $ &$ [0.2,1]$ & $ [3.70,14.69]$&$225 $  \\ 
\hline
\end{tabular}}
\caption{\footnotesize{Characteristics of the different clusters samples.}}
\label{tab:cat}
\end{center}
\end{table}

We compare results on the $\rm NC^{\rm tSZ} + \rm BAO \, + \, (1-\mathcal{B}) \, + \alpha_b \, + \, \beta_b$ dataset combination for the different samples. The constraints are reported in Table~\ref{tab:mz_cat}, where for comparison purposes we report once more the constraints for the entire PSZ2 cosmo catalogue. 

Focussing on the $\alpha_b$ and $\beta_b$ parameters, we see how results change considering the different samples. In particular, changing the signal-to-noise ratio threshold shifts the $\alpha_b$ parameter from being completely consistent with 0 ($\alpha_b = 0.10 \pm 0.10$ at $q_{\rm min} =6$) to providing a strong hint of mass dependence ($\alpha_b = 0.38_{-0.03}^{+0.12}$ at $q_{\rm min} =8.5$), pointing the $(1-b)_{\rm var}$ quantity to increase with mass. On the contrary, the $\beta _b$ parameter is shifted towards lower values, completely in agreement with 0, from $\beta_b = 0.24_{-0.18}^{+0.24}$ at $q_{\rm min} =6$ to $\beta_b = -0.24 _{-0.38}^{+0.36}$ at $q_{\rm min} =8.5$.

When considering the different redshift ranges, on the one hand $\alpha_b$ remains consistent with 0, with $\alpha_b = 0.10 \pm 0.10$ for $z$ in $[0,1]$ and $\alpha_b = 0.04 \pm 0.10$ for $z$ in $[0.2,1]$. On the other hand, $\beta_b$ changes from $\beta_b = 0.24_{-0.18}^{+0.24}$ for $z$ in $[0,1]$ to $\beta_b = -0.09_{-0.17}^{+0.23}$ for $z$ in $[0.2,1]$.

\begin{figure}[!h]
\centering
\includegraphics[scale=0.48]{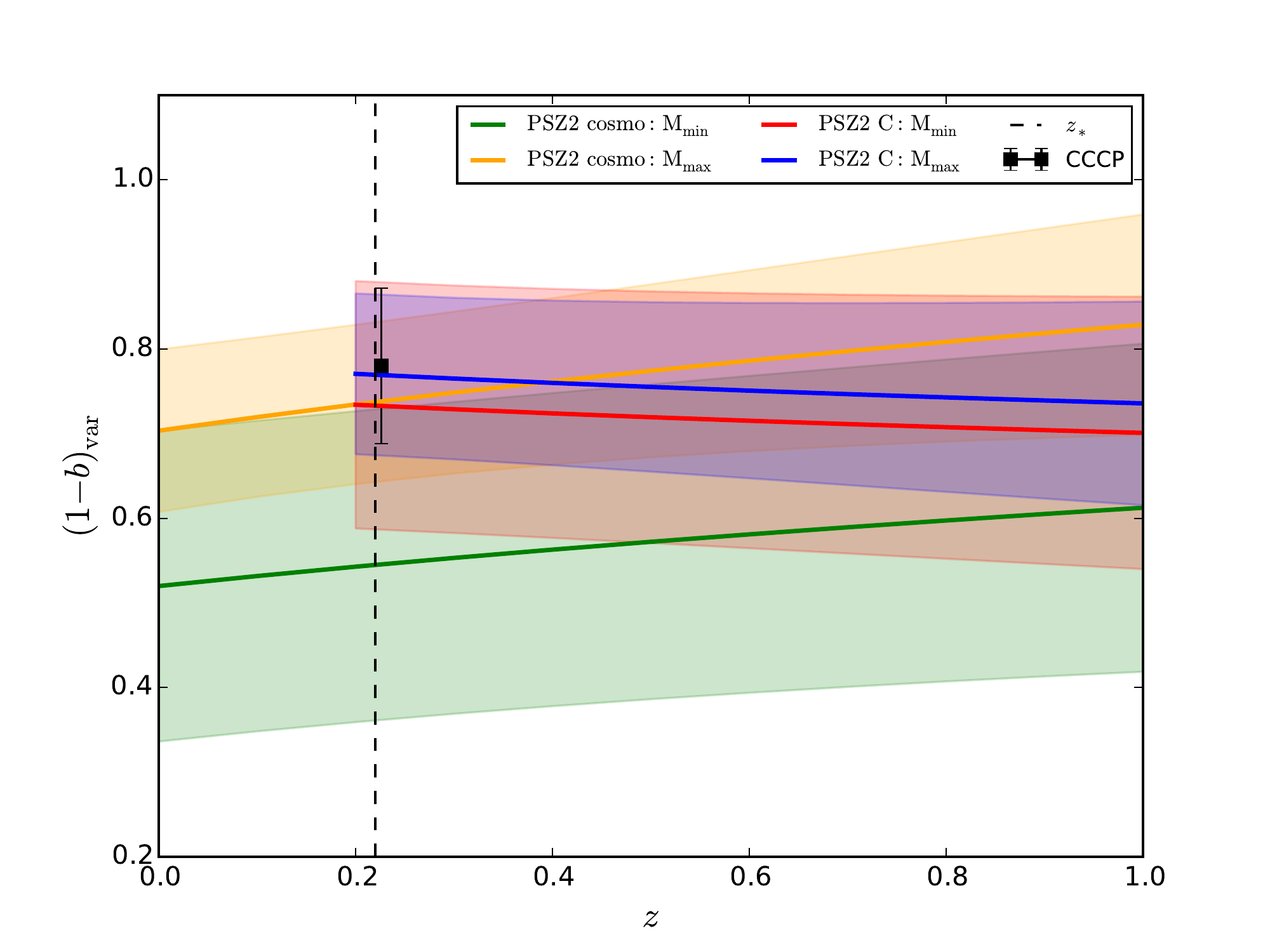}
\caption{\footnotesize{Redshift variation of $(1-b)_{\rm var}$ at fixed values of mass. We compare results for the complete PSZ2 cosmo sample and the PSZ2 C sample, obtained selecting $z\gtrsim 0.2$ clusters, in both cases adopting CCCP calibration.}}
\label{fig:bias_var_zmin}
\end{figure}

Focussing in particular on the redshift dependence, we show in Fig.~\ref{fig:bias_var_zmin} the $(1-b)_{\rm var}$ variation for the PSZ2 cosmo and PSZ2 C samples. We can clearly see that neglecting the low redshift clusters induces an opposite redshift dependence, leading to a decreasing trend of $(1-b)_{\rm var}$ for higher redshift.

The changes on the $\alpha_b$ and $\beta_b$ parameters also provide a shift in the results for the amplitude $(1-\mathcal{B})$. Nevertheless, the constraints on $\Omega_m$ and $\sigma_8$ remain consistent for the different samples (see again Tab.~\ref{tab:mz_cat}). We report a triangular plot in Appendix \ref{sec:app}, showing the constraints and correlations for the cosmological and mass bias parameters for the different samples (Fig.~\ref{fig:SELcomp}).
We therefore conclude that constraints on the mass-redshift dependence for the mass bias are completely dependent on the sample that we are considering.

\begin{table*}
\begin{center}
\scalebox{0.9}{
\begin{tabular}{c|c|c|c|c|c}
\hline
\hline
\morehorsp
 $\rm NC^{\rm tSZ} + \rm BAO \, + \, (1-\mathcal{B}) \, + \alpha_b \, + \, \beta_b$& $\Omega _m$ & $\sigma_8$ & $(1-\mathcal{B})$ & $\alpha _b $ & $\beta _b$\\
\hline
\morehorsp
PSZ2 cosmo, CCCP &$0.374 _{-0.035}^{+0.065} $ &$0.752 _{-0.040}^{+0.033} $ & $0.66_{-0.14}^{+0.10}$ &$ 0.10 \pm 0.10$ &$0.24_{-0.18}^{+0.24} $\\
\hline
\morehorsp
PSZ2 A, CCCP &$0.380_{-0.037}^{+0.062} $&$0.772_{-0.038}^{+0.031} $ &$0.51 \pm 0.10 $ &$0.29 \pm 0.11 $ &$0.01 _{-0.24}^{+0.31} $\\
\hline
\morehorsp
PSZ2 B, CCCP &$0.364 _{-0.040}^{+0.054} $&$0.770 _{-0.039}^{+0.032} $ &$0.444 _{-0.084}^{+0.057} $ &$0.38_{-0.03}^{+0.12} $ &$-0.22 _{-0.32}^{+0.39} $\\
\hline
\morehorsp
PSZ2 C, CCCP &$0.378 _{-0.036}^{+0.048} $&$0.740 _{-0.044}^{+0.034} $ &$0.74 _{-0.15}^{+0.12} $ &$0.04 \pm 0.10 $ &$-0.09_{-0.17}^{+0.23} $\\
\hline
\hline
\end{tabular}}
\caption{\footnotesize{$68\%$ c.l. constraints for cosmological and mass bias parameters, for the different dataset combinations.}}
\label{tab:mz_cat}
\end{center}
\end{table*}

\section{Discussion}\label{sec:discussion}

Calibration of the scaling relations, the mass-redshift evolution, and the possible departure from the assumptions of hydrostatic equilibrium and self similarity have raised much interest in the community. Focussing in particular on the mass bias, the Planck collaboration has provided such a broad galaxy cluster catalogue that many authors have contributed with evaluations of cluster masses and with possible estimation of mass and redshift dependence for the mass bias, selecting different sub-samples of the catalogue, see for example \cite{vonderLinden:2014haa,Hoekstra:2015gda,Okabe:2015faa,Sereno:2015jda,Smith:2015qhs,Penna-Lima:2016tvo,Sereno:2017zcn} and a collection of measurements in \cite{Salvati:2017rsn}. Given how the discrepancy on cosmological parameters between tSZ probes and CMB data has been partially solved thanks to new measurements of the optical depth, the correct description of the scaling relations remains one of the open issues when using galaxy clusters as a cosmological probe, with the mass bias being one of the largest source of uncertainties (see e.g. the recent review from \cite{Pratt:2019cnf} for a full description of the impact of the mass bias on the cosmological constraints).

The mass bias is generally introduced to quantify any departures from the assumption of hydrostatic equilibrium when evaluating clusters masses with X-rays and tSZ observations. 
The uncertainties in the evaluation of the mass bias can be related to the description of cluster physics (e.g. baryonic effects, processes leading to non-thermal pressure contributions, magnetic field) or to experimental uncertainties in tSZ and X-rays measurements.
Hydrodynamical simulations provide an alternative estimation of this quantity, $(1-b) \sim 0.8$, with the main uncertainties being related to the identification and evaluation of the different processes leading to non-thermal pressure contributions, see for example \cite{Biffi:2016yto} and references in \cite{2014A&A...571A..20P}.

Given this complex scenario, 
we decided to focus our analysis on the mass bias. We considered the scaling relations provided from the Planck collaboration (and reported in Eqs.~\eqref{eq:Y500} and \eqref{eq:theta500}). We followed the general assumptions on self-similarity, hydrostatic equlibrium and mass-redshift evolution for the scaling relations and encode all possible departures from these assumptions in the mass bias.
We propose a simple and empirical mass and redshift parametrisation for the mass bias. In discussing the results, we focus both on the constraints on the mass bias itself and on the effects on the CMB-tSZ discrepancy. Since the stronger constraining power of CMB data does not allow to have any mass or redshift variation, for the first part of the discussion we focus only on results from tSZ+BAO data.

In the Section \ref{sec:results}, we showed that the constraints on the $\alpha_b$ and $\beta_b$ parameters from the $C_{\ell}^{\rm tSZ} + \rm NC^{\rm tSZ} + \rm BAO$ combination are marginally consistent with 0.  Nevertheless, when evaluating the total $(1-b)_{\rm var}$ quantity, we clearly see a hint for a mass-redshift evolution. In particular, we find $(1-b)_{\rm var}$ to be slightly increasing with redshift. This result is partly enhanced when considering only $\rm NC^{\rm tSZ} + \rm BAO$.

In order to provide a more complete analysis, we compared results from different mass bias calibrations, when considering the $\rm NC^{\rm tSZ} + \rm BAO$ combination. In particular, we considered results from another WL calibration (the WtG analysis) and from a cosmological hydrodynamical simulation. Regarding the WL evaluations, we decided to compare the CCCP and WtG analysis, to highlight how lower and higher values of the mass bias affect the constraints on $\Omega_m$ and especially $\sigma_8$. When comparing these results it is important to recall how the calibrations are estimated. For all the details, we refer to the single analysis (\citealp{Hoekstra:2015gda,vonderLinden:2014haa}). We note here that the different works consider diverse sub-samples of the entire Planck cosmological catalogue, with different number of clusters, mass and redshift range and yet with some overlapping of objects between the different selections. Furthermore, the lensing mass extraction methodologies differ from one evaluation to another. 
Regarding hydrodynamical simulations, we consider results from \cite{Biffi:2016yto}, which evaluates mass bias for a sample of simulated clusters at redshift $z=0$. We stress that even comparing different evaluations for the mass bias (based on different approaches and considered cluster samples), we find the same slight increasing trend of $(1-b)_{\rm var}$ towards high redshift. Indeed, apart from different values of the calibrations themselves, 
the shape of $(1-b)_{\rm var}$ is nearly independent of the pivots at which these calibrations are applied. 
We compared the results obtained with the different calibrations with the case when we applied a flat prior on the total mass bias. By diminishing the range for $(1-b)_{\rm var}$, we also reduce the allowed values for $\alpha_b$ and $\beta_b$, therefore removing any hint of redshift and mass variation. Nevertheless, we stress that even without considering any external calibration, the preferred value for the mass bias is given by $(1-\mathcal{B}) = 0.756_{-0.083}^{+0.056}$.

The slightly increasing trend of $(1-b)_{\rm var}$ with redshift is also confirmed when we divide the redshift range of the PSZ2 cosmo sample in three bins and evaluate the mass bias in each bin. Indeed, we find the $(1-b)_i$ parameters moving towards higher values for higher redshift.

The CCCP and WtG analysis provide results regarding a mass dependence for the mass bias, while not discussing a possible redshift dependence. Both WtG and CCCP agree in finding a modest evidence for the mass dependence, having $M_{\rm Planck}\propto M_{\rm WtG}^{0.76_{-0.20}^{+0.39}}$ and $M_{\rm Planck} \propto M_{\rm CCCP}^{0.64 \pm 0.017}$. 

In the collection of CoMaLit papers \citep{Sereno:2014pfa,Sereno:2014qfa,Sereno:2014aea,Sereno:2015jda,Sereno:2016cut} the authors provide an extensive discussion on measurements and calibrations for the scaling relations, analysing different wavelengths results. In particular, in \cite{Sereno:2016cut}, the authors focus on the analysis of a possible mass and redshift dependence for the mass bias, comparing WL and tSZ estimated masses for a sub-sample of the Planck cosmological catalogue of 135 galaxy clusters. 
Their analysis provides a mass bias that is nearly mass independent but increases with redshift, implying the cluster masses to be strongly underestimated at higher redshift. We find consistent results when considering the PSZ2 C sample, cutting at $z_{\rm min} =0.2$. 

In \cite{Smith:2015qhs}, the authors evaluate masses for a sub-sample of 44 Planck clusters in the redshift range $z=[0.15,0.3]$, finding the value $(1-b) = 0.95 \pm 0.04$. When comparing this value with WtG and CCCP analyses, they find that the different results are due not only to different methods of mass calculation, but also to the different redshift ranges. 
Indeed, when splitting WtG and CCCP redshift ranges in two bins ($z<0.3$ and $z>0.3$) and re-evaluating clusters masses, they find a general agreement on the mass bias for the lower redshift bin, highlighting a decreasing trend for $(1-b)$ with respect to redshift. We again find the same trend when considering the PSZ2 C sample, with $z_{\rm min} =0.2$

There are also attempts to constrain the mass-redshift dependence of the mass bias based on tSZ power spectrum. In \cite{Makiya:2018pda}, the authors provide a joint analysis of Planck tSZ power spectrum and the number density fluctuations of galaxies in the Two Micron All Sky Survey (2MASS) redshift survey (2MRS). Despite showing the 2MRS-tSZ cross-power spectrum to be more sensitive to less massive haloes than the tSZ autopower spectrum, they are not able to strongly constrain a mass variation for the mass bias.

To summarize, we stress the difficulty in comparing our results and the different analyses. We have shown how the mass and redshift dependence may depend on the sample definition, providing a hint of different behaviour for low redshift ($z<0.2$) and high redshift ($z>0.2$) clusters. In particular, we highlight that different cuts in redshift (and signal-to-noise ratio) also change the considered mass range. In Tables~\ref{tab:zbins} and \ref{tab:cat} we report the mass ranges for the different sub-samples of the PSZ2 cosmo sample. The lower redshift bin, up to $z=0.2$, is the one containing the lower masses objects. It is possible that our results on the mass bias evolution are affected by this asymmetric mass distribution of clusters in the entire redshift range. 
Furthermore, the large amount of low-mass clusters in $z<0.2$ could also explain the low value for $(1-b)$ that we find in that redshift interval, having $(1-b) \sim 0.6$.

Therefore, in order to improve these results and provide a more realistic description of the mass-redshift evolution of the mass bias, it is necessary to have access to WL calibrations based on wider sub-samples, more representative of the cluster population that we consider for the cosmological analysis.\\

Next, we focus on the evaluation of the mass bias from CMB and tSZ data combination. We discuss above how the latest results on the optical depth from the Planck collaboration have significantly reduced the discrepancy on the $\sigma _8$ parameter between tSZ and CMB data, up to $1.5 \, \sigma$ and, as a consequence, the one on the mass bias. Nevertheless, CMB primary anisotropies prefer values of the mass bias that are still higher than evaluations from simulations, WL and Xrays analysis (see e.g. \cite{Maughan:2015fxf}), having $(1-b) \simeq 0.6$.

We have also compared the tSZ+CMB results with alternative estimations of the mass bias. We consider the analysis presented in \citep{Ettori2019,Eckert2019}. The authors provide constraints on the mass bias through the evaluation of non-thermal pressure contributions. The authors analyse how physical quantities that describe the intracluster medium deviate from the pure gravitational collapse model and therefore from the condition of hydrostatic equilibrium. In particular, they focus on the evaluation of the gas fraction for 12 local massive clusters in the Planck cosmological catalogue, from which they obtain the mass bias constraints $(1-b) = 0.85 \pm 0.05$. Furthermore, they stress how constant values of mass bias $(1-b)\sim 0.6$ result in a too low value for the universal gas fraction, being rejected at almost $4\, \sigma$.

In previous sections we analysed if the mass-redshift parametrisation proposed in Eq.~\ref{eq:1mb_par} can help in shifting the mass bias constraints towards higher values, for the CMB+tSZ dataset combination. However, when we apply the `$(1-\mathcal{B}) + \alpha_b + \beta_b$' parametrisation, we find results that are consistent with the standard scenario, having $(1-b)_{\rm var} \simeq 0.6$. 
Furthermore, as shown in Fig.~\ref{fig:mz_bvar_Mz}, the variation of $(1-b)_{\rm var}$ for tSZ data allows only for partial consistency between the two datasets combinations.
We conclude, therefore, that a mass-redshift parametrisation of the mass bias does not help in completely reconciling results from CMB and tSZ data, still providing constraints that are not in agreement with other astrophysical evaluations.

\section{Conclusions}\label{sec:conclusions}
We analysed a possible dependence for the mass bias with respect to mass and redshift, considering galaxy clusters observed through the tSZ effect by the Planck satellite and combining galaxy clusters number counts with estimation of tSZ power spectrum. We compared and combined tSZ data with the latest CMB results from the Planck satellite. 

We considered an empirical mass-redshift parametrisation and perform several tests in order to check the consistency of our results. 
When considering tSZ+BAO data, we find a modest hint for redshift dependence, leading the $(1-b)_{\rm var}$ quantity to increase with redshift. We tested that these results do not depend on the choice of the mass bias calibration, by comparing the effect of WL and hydrodynamical calibrations. As a further test, we divided the redshift range for the PSZ2 cosmo sample in 3 bins and analyse the change in the $(1-b)_i$ parameters. We find the same hint for redshift dependence, having the $(1-b)_i$ parameters increasing towards higher redshift. We also analysed the case in which we do not consider any external calibration. We highlight that even in this scenario, the preferred value for $(1-b)$ from tSZ galaxy clusters is equal to $0.756 _{-0.083}^{+0.056}$.

We stress, however, that these results on the mass-redshift variation depend on the selected cluster sample. Indeed we compared results from samples obtained considering different signal-to-noise ratio thresholds and redshift range. In general, we find the constraints on $\alpha_b$ and $\beta_b$ (and therefore the mass and redshift dependence) to change for the different samples. It is therefore difficult to fully compare these results with other analyses available in literature.

We analysed results for the complete tSZ+CMB data combination. We find statistically consistent results on the $(1-b) _{\rm var}$ quantity. Nevertheless, we stress that CMB data still pushes towards high values of the mass bias (i.e. low values of $(1-b)$), providing results not in agreement with numerical simulations and WL estimations.
We conclude, therefore, that a mass-redshift variation, at the current level of precision and accuracy of observations, does not solve the discrepancy on the mass bias calibration between CMB and large scale structure evaluations.

\begin{acknowledgements}
      LS acknowledges support from the postdoctoral grant from Centre National d'\'Etudes Spatiales (CNES). AB acknowledges support from NSERC's Discovery Grant programme. 
The authors thank Veronica Biffi for providing useful information on the simulated cluster sample used in \cite{Biffi:2016yto}. The authors also thank the referee for helping to improve the analysis.
Based on observations obtained with Planck (http://www.esa.int/Planck), an ESA science mission with instruments and contributions directly funded by ESA Member States, NASA, and Canada. This project made use of the SZ-Cluster Database (http://szcluster-db.ias.u-psud.fr) operated by the Integrated Data and Operation Centre (IDOC) at the Institut d’Astrophysique Spatiale (IAS) under contract with CNES and CNRS. This project has received funding from the European Research Council (ERC) under the European Union's Horizon 2020 research and innovation programme grant agreement ERC-2015-AdG 695561.
      
\end{acknowledgements}

\begin{appendix}

\section{Cosmological and mass bias parameters constraints}\label{sec:app}

\begin{figure*}[!h]
\centering
\includegraphics[scale=0.99]{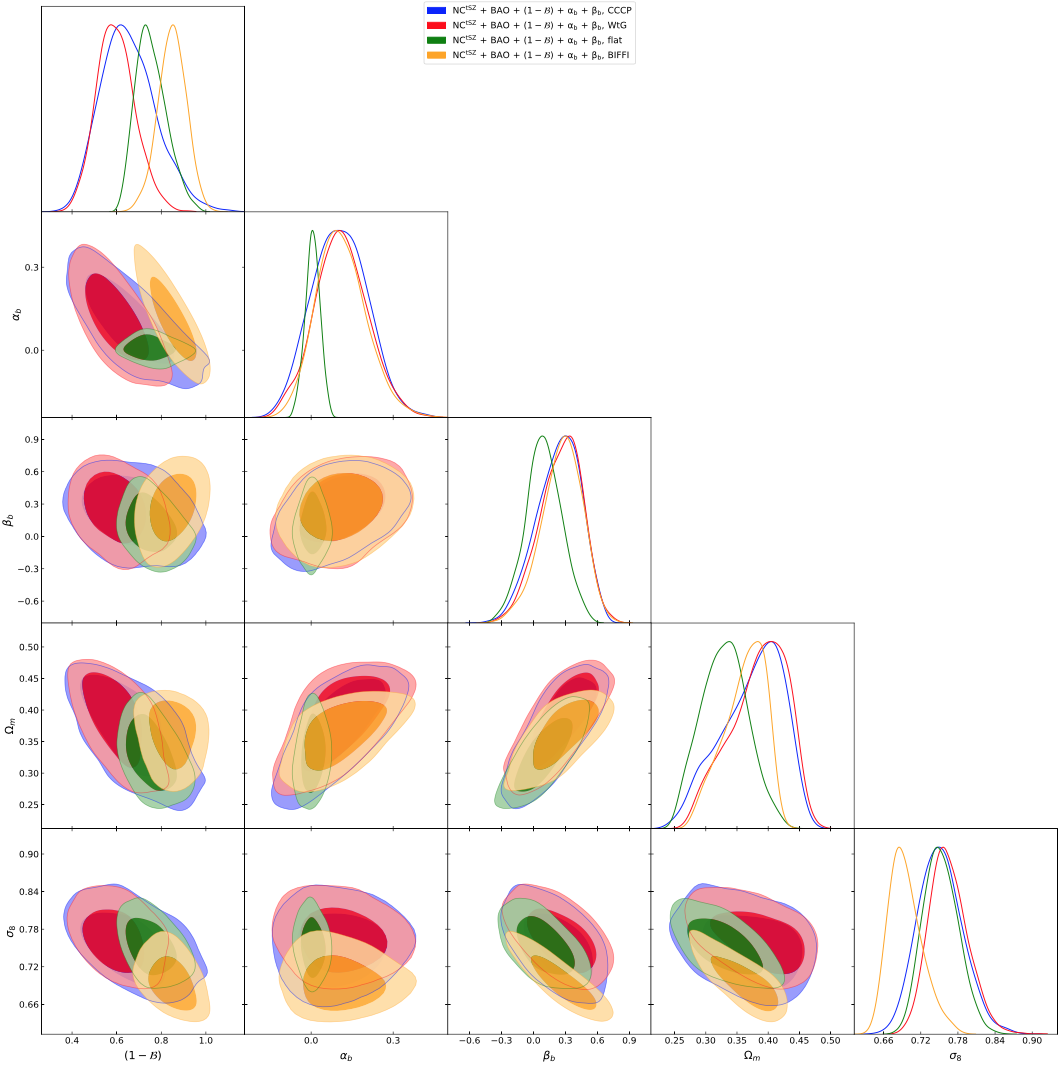}
\caption{\footnotesize{Correlation between cosmological and mass bias parameters, for the $\rm NC^{\rm tSZ} + \rm BAO \, + \, (1-\mathcal{B}) \, + \alpha_b \, + \, \beta_b$ combination. In blue we show results for the CCCP calibration (baseline), in red for the WtG calibration, in orange for BIFFI and in green results when using a flat prior, $(1-b)_{\rm var} = [0.6,1.0]$.}}
\label{fig:biasCAL}
\end{figure*}

\begin{figure*}[!h]
\centering
\includegraphics[scale=0.94]{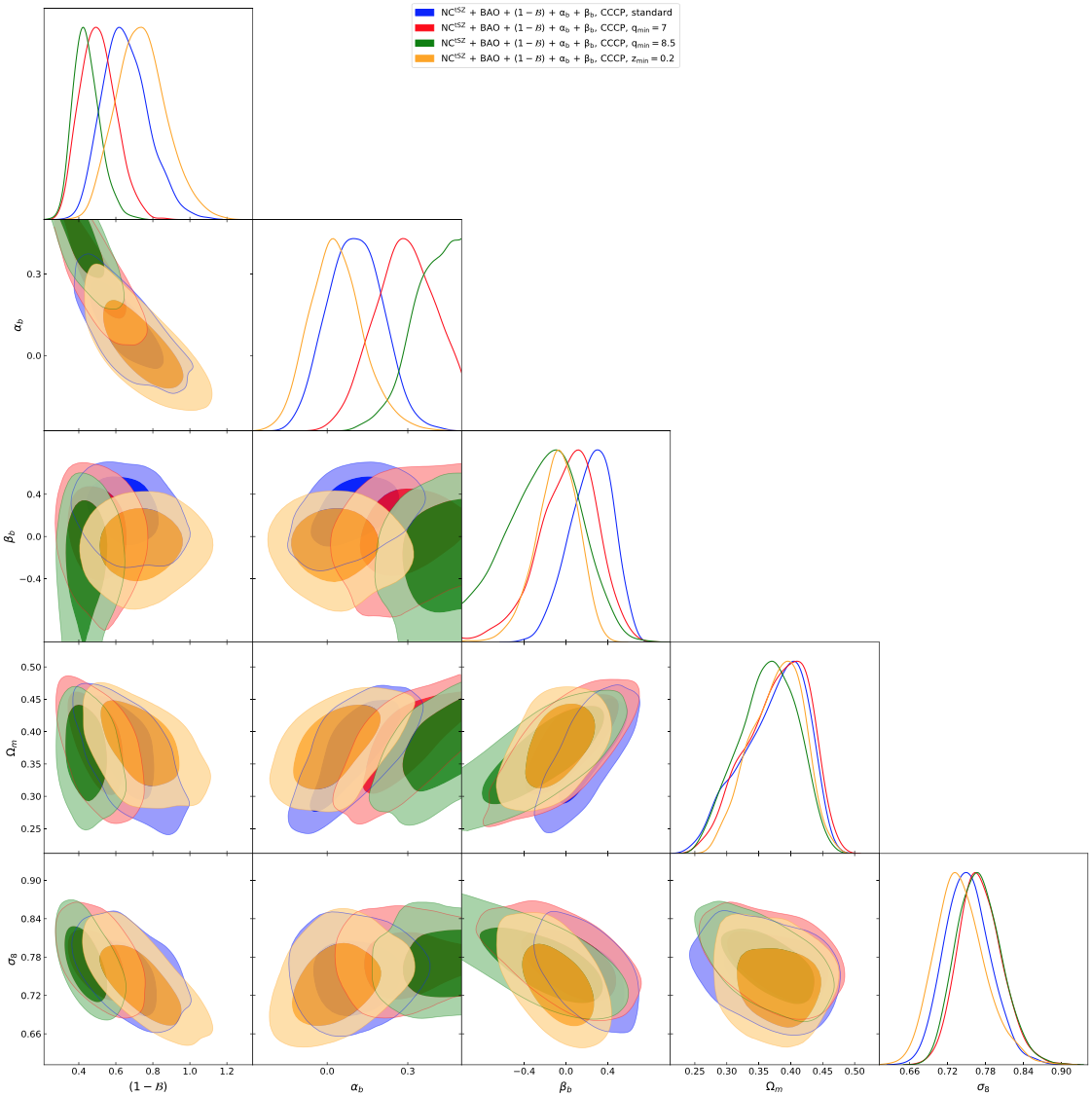}
\caption{\footnotesize{Correlation between cosmological and mass bias parameters, for the $\rm NC^{\rm tSZ} + \rm BAO \, + \, (1-\mathcal{B}) \, + \alpha_b \, + \, \beta_b$ combination. In blue we show results for PSZ2 cosmo sample ($q_{\rm min} = 6$), in red for the PSZA sample ($q_{\rm min} = 7$), in green for the PSZB sample ($q_{\rm min} = 8.5$) and in orange for the PSZC sample ($z_{\rm min} = 0.2$).}}
\label{fig:SELcomp}
\end{figure*}

We show in the triangular plots in Fig.~\ref{fig:biasCAL} and \ref{fig:SELcomp} the one-dimensional and two-dimensional probability distributions for the cosmological ($\Omega_m$ and $\sigma_8$) and mass bias parameters. We report constraints for the different mass bias calibrations and sub-samples of the entire PSZ2 cosmo catalogue, as described in Sect.~\ref{sec:tests}.

\end{appendix}

\bibliographystyle{aa} 
\bibliography{references}

\end{document}